# Development of machine-learned interatomic potentials to predict structure, transport, and reactivity in platinum-based fuel cells

Kamron Fazel,[1, a)] Sam Brown,[2, a)] Jacob Clary,[3] Pritom Bose,[4] Nima Karimitari,[5] Amalie L. Frischknecht,[6] Ravishankar Sundararaman,[1, b)] and Derek Vigil-Fowler[3, c)]

[1)] *Materials Science & Engineering, Rensselaer Polytechnic Institute, Troy, NY 12180, USA*
[2)] *Department of Chemistry and Biochemistry, New Mexico State University, Las Cruces, NM 88003, USA*
[3)] *Materials, Chemical, and Computational Science Directorate, National Laboratory of the Rockies, Golden, CO 80401, USA*
[4)] *Mechanical, Aerospace & Nuclear Engineering, Rensselaer Polytechnic Institute, Troy, NY 12180, USA*
[5)] *Department of Chemistry and Biochemistry, University of South Carolina, Columbia, SC 29208, USA*
[6)] *Center for Integrated Nanotechnologies, Sandia National Laboratories, Albuquerque, New Mexico 87185, USA*

Machine-learned interatomic potentials (MLIPs) have rapidly progressed in accuracy, speed, and data efficiency in recent years. However, training robust MLIPs in multicomponent systems still remains a challenge. In this work, we train a MLIP to describe hydrated Nafion ionomers and platinum catalysts, which are important components of fuel cells, by constructing a diverse training set to describe the bulk polymer and interfacial catalyst-polymer interactions well. We find that active learning improves the initial dataset little in terms of reducing uncertainty and error, pointing towards a need for more effective methods to efficiently explore the relevant interactions in complex, multicomponent systems. We use our trained MLIP to study the properties of the platinum-Nafion system, including polymer structure, proton mobility in a bulk Nafion polymer and near a platinum-Nafion interface, and reactions near and far from the interface, finding excellent results for structure and reactions contained within our training set. Transport seems well described, with both vehicular transport and Grotthuss hopping captured, although converged calculations of diffusivities were not computed because they require calculations of tens of nanoseconds that are challenging with current state-of-the-art MLIPs. The combined insights that this model provides can be leveraged to optimize fuel cell performance, and the approach can be applied to other chemical processes and devices where structure, transport, and reactivity all contribute to overall observed performance.

## I. INTRODUCTION

Hydrogen fuel cells have been developed and deployed for use in transportation and stationary applications[57] for decades, with significant efforts under way to make them more durable and efficient for heavy-duty applications[14]. They could also help meet the energy needs of rapidly expanding data centers[36]. Improvements in the efficiency and cost of fuel cells could thus have significant positive economic impacts, but fuel cells are complex devices that require significant testing and insight at multiple length and time scales for optimization.

Mainstream fuel-cell technology uses proton-exchange membranes (PEM) made from Nafion and platinum (Pt)-based catalysts for the reaction at both the anode and cathode[17]. The hydrogen oxidation[52] and oxygen reduction reactions (ORR)[22] generate and consume protons, respectively, and the protons are transported from the anode to the cathode via water domains[45] formed in the hydrated Nafion membrane. Among the challenges of optimizing PEM fuel cells is the coupled behavior of proton transport via the Nafion polymer, the reactivity of the Pt catalyst, and their interface.[17,27] Many atomistic computational studies using classical molecular dynamics (MD) simulations, density functional theory (DFT) and *ab initio* MD (AIMD), or MLIPs have been carried out to characterize PEM- and platinum-based fuel cells to study either transport or reactivity, although rarely are both treated within a single methodology.

Classical MD studies frequently predict water and hydronium dynamics, but usually do not account for bond-forming and breaking events given the difficulty of developing reactive classical force fields.[8,9,25,29,35,37,49] The treatment of reaction chemistry and other quantum mechanical phenomena is more commonly performed using DFT or higher accuracy theories. However, as DFT-based calculations are limited to shorter length and time scales, they are sometimes combined with classical MD simulations to get a more complete picture of how different components of a fuel cell interact. For example, Brunello, et al., used DFT to understand Pt nanoparticle formation and classical MD to model the interactions of these nanoparticles with polymer electrolytes.[6]

MLIPs have the potential to enable high-accuracy studies of much larger systems than are possible to treat with AIMD, bridging the gap between the computational efficiency of classical MD and the accuracy of AIMD. However, MLIPs can be challenging to train for polymers due to the large configuration spaces involved. Techniques such as Gaussian process regression and active learning have been used to improve MLIP accuracy by iteratively adding training data in areas of high model uncertainty.[7] Indeed, these approaches have been used to train models that model physical properties of bulk

---

[a)]These authors contributed equally to this work.
[b)]Electronic mail: sundar@rpi.edu
[c)]Electronic mail: derek.vigil-fowler@nrel.gov

Table I. Summary of the initial training set structures that were run using AIMD in the NVT ensemble prior to active learning. Every $10^{th}$ frame was selected for training. The middle three rows are collectively referred to in this work as Nafion-water systems. The Nafion chain used in the last two rows was Nafion[x=1,y=2,z=1] and shown in Figure S2.

| System | $N_{chains}$ | $\lambda$ ($H_2O$ per $SO_3$ group or chain) | Polymer rotations | Linear strain (%) | Strain direction | Number of fs per case | Total frames for system type |
|---|---|---|---|---|---|---|---|
| Water | - | - | - | -10, -5, -2, 0, 1, 2, 5 | xyz | 2000 | 14000 |
| $CF_3$-$3CF_2$-$CF_3$ | 4 | 6, 12, 18, 24 | - | -15, -10, -5, -2, 0 | xyz | 2000 | 40000 |
| $HSO_3$-$2CF_2$-O-$CF_3$ | 4 | 6, 12, 18, 24 | - | -15, -10, -5, -2, 0 | xyz | 2000 | 40000 |
| Nafion[x=1,y=2,z=1] | 1 | 6, 12, 18, 24 | - | -15, -10, -5, -2, 0 | xyz | 2000 | 40000 |
| (4×4) Pt(111)+Nafion | 1 | 6, 12, 18, 24 | 10 | -15, -10, -5, -2, 0 | z | 500 | 100000 |

polymers[24] and predict proton transport in Nafion over much longer timescales than AIMD with cell sizes comparable to those used in AIMD.[28] Various advancements in MLIPs specifically for the study of polymers have been described in greater detail in a recent review[41]. However, it is unclear how well such MLIPs perform for heterogeneous environments that include metal catalyst-polymer interfaces.

Recent advances in neural network-based MLIP architectures have yielded roughly an order of magnitude improvement in prediction errors using orders of magnitude less training data.[34] One prominent example is the multi-atomic cluster expansion (MACE) approach[3,5]. In the MACE architecture, the final energetic contribution of each atom is expressed as a learnable function of many-body atom-centered features describing each atom's local environment. MACE models have been used to study a variety of systems with high accuracy, including perovskites[30], drug-like molecules[20], liquid water[32], and disordered crystals[32]. Further, large collaborations have trained MACE foundational models using the Materials Project database (MACE MP-0 model)[4] and organic datasets (MACE OFF model)[33] that exhibit good accuracy across diverse test applications. These foundational models leverage large dataset training efforts, but can incur a larger computational cost than custom MLIPs trained from scratch for specific systems.

In this work, we trained a MACE MLIP to predict polymer structure, proton mobility in the bulk and at the interface, and proton-transfer and polymer dissociation reactivity in a composite Pt-Nafion-water system. We did this by constructing a diverse training set that describes the bulk polymer and catalyst-polymer interface well, including volume compressions and different orientations of polymer chains on the platinum surface to capture different binding motifs. We also show the effect of active learning on model performance.

## II. METHODS

### A. MLIP training

The training set was constructed to describe the complexity of the catalyst-polymer interface in a fuel cell. We ran AIMD trajectories for water, Nafion-water, and Pt-Nafion-water systems initialized with different hydration levels and strains to span the expected range of hydrations and densities known to be present in hydrated Nafion membranes. Figure S1 in SI Section I summarizes the densities for systems containing Nafion fragments or a Nafion chain. The structure space of Pt-Nafion interactions was sampled by rotating a 1125 g/mol Nafion chain (Figure S2) in 40° steps about the x, y, and z axes and selecting 10 (Figure S3) that represented a diverse set of Pt-Nafion interactions. Table I gives a full description of the training data for our model. Structure diversity was assessed by using the ASAP package[10] to generate SOAP descriptors[2] of the structures and down-selecting using the quota sampling approach in the iSIM package[42]. The AIMD simulations were performed with the open-source JDFTx software[54] using the LibXC[43] r2SCAN meta-generalized-gradient approximation functional[19], a plane-wave basis with a 30 hartree wavefunction kinetic energy cutoff, and SG15 norm-conserving pseudopotentials[23,48]. Additional details are in SI Section IV. Every tenth frame was included in the training dataset to reduce correlated structure data.

Model training using the MACE architecture was performed in an active learning workflow, utilizing a committee of 3 models that were all trained on the same dataset with different training randomization seeds. The training parameters used can be found in SI Section II. During each iteration, one of the three models was used to produce 24 separate 1 ps NVT trajectories, where each of the trajectories was initialized using the last snapshot of a corresponding training AIMD trajectory as the starting point. The NVT simulations were done using a Langevin thermostat at 300 K. More information on our active learning approach can be found in SI Section III, including the number of flagged frames for each sys-tem, relative force magnitude deviation statistics, and energy deviations shown in Tables S1-S3 for the 3 active learning iterations. The tables show that most of the force deviation among models arises from the platinum-containing systems, but the energy deviation of the water systems and Nafion-water systems are much higher than the Pt-containing systems. The lower energy deviation in Pt-containing systems can likely be attributed to the higher number of frames in the training data of these systems, while the increased force error in the Pt-containing systems is likely due the difficulty in learning Pt-Nafion-water interface interactions.

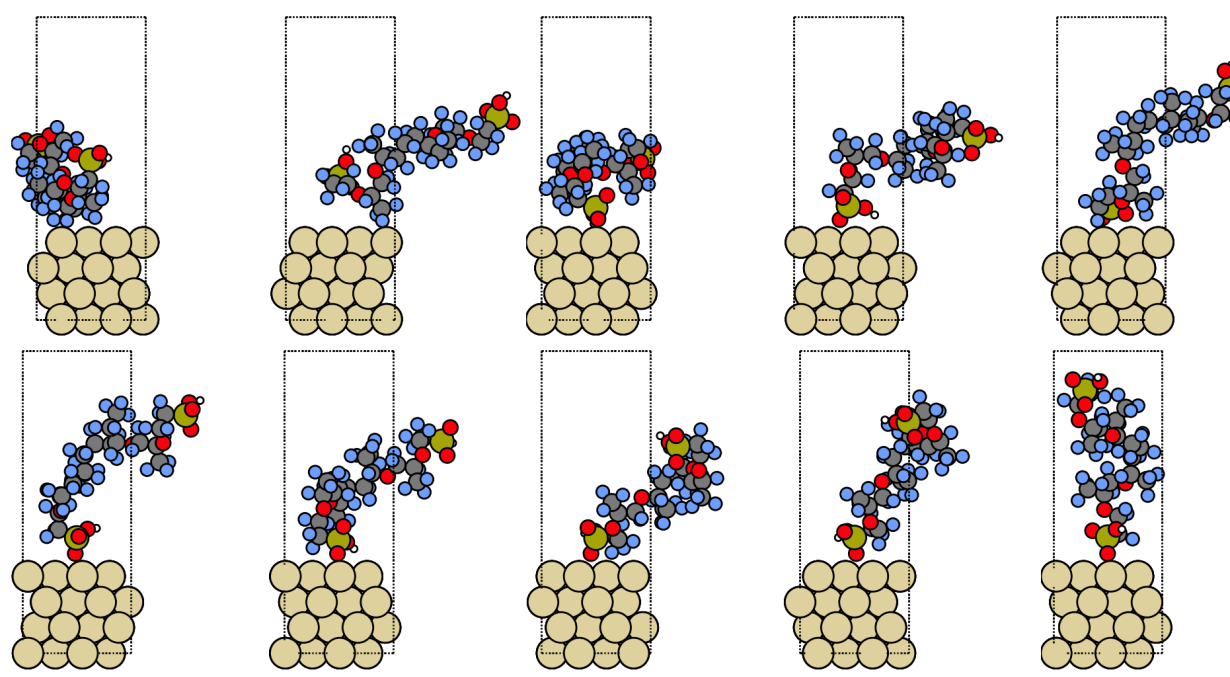

Figure 1. Rotations of the Nafion[x=1,y=2,z=1] fragment in Figure S2 selected to sample the Pt-Nafion interaction space. We note that the polymer chains wrap at the periodic supercell boundary. Platinum, hydrogen, oxygen, carbon, sulfur, and fluorine atoms are represented as light brown, white, red, gray, gold, and blue spheres, respectively.

Table II shows the validation errors for our MLIPs at each active learning iteration. We found that active learning using the above approach did not improve our model's force and energy prediction performance, with either force or energy-prediction accuracy worsening in our active learning iterations 1 and 2 relative to the initial model. Tables S1-S3 also show that our model uncertainty is improved by active learning, but that there are diminishing returns after the first iteration. This suggests that our initial training set likely already contained the relevant atomic interactions and environments, and active learning did not effectively add further sampling of this complex, multicomponent system. As a result, the frames added via active learning were not included in the final model used to generate the results in this work. Alternative approaches that can more precisely target particular element environments for further model improvement would be highly desirable, particularly for diverse, multicomponent systems such as these. Small model improvements due to active learning over random sampling in a high-dimensional problem space has also been reported for predicting the glass transition temperature of a polymer set.[31]

To understand which parts of the training data set are the most important for achieving high accuracy in energy and force predictions, we performed an ablation test in which we removed different parts of the training set: half of the Nafion rotations on platinum, the strained systems, the Nafion-water systems, or all but the -15% strained systems. From the results in S5, we see that data from the strained systems is by far the most important for achieving high accuracy energy and force predictions, and most of the benefit comes from the highest compression (-15% strain) systems. The Nafion rotations on platinum primarily give extra accuracy in the forces, while the Nafion-water systems give improvements in both the energy and force predictions. While it is possible that similar accuracy could be achieved for the results calculated below with less data in the training, we leave the study of this possibility to a future investigation.

Table II. Training force (meV/atom) and energy (meV/Å) validation errors of active learning iterations.

| Iteration | $\langle$RMSE($E$)$\rangle$ (meV/atom) | $\langle$RMSE($F$)$\rangle$ (meV/Å) | Training Configs |
| --- | --- | --- | --- |
| Initial | 5.03 | 50.83 | 23868 |
| Iteration 1 | 4.87 | 50.87 | 24541 |
| Iteration 2 | 5.47 | 50.80 | 24928 |

## B. Model application

As this study aims to create a model capable of predicting Nafion structure, reaction energetics, and proton diffusion both near to and far from the Pt surface, the test systems described below were created to be large enough to enable such insights while still being computationally tractable with existing computational resources. These systems were equilibrated using classical MD simulations in order to provide good starting configurations for simulations with the trained MACE potentials.

All systems consisted of 9 Nafion chains, each with 6 repeat units consisting of 14 $CF_2$ groups and a side chain branch. The Nafion chains had an equivalent weight of 1143 g/mol per repeat unit. The sulfonate groups were assumed to be completely disassociated. Systems were built with three water contents, $\lambda$, of 9, 12, and 15 waters per sulfonate group. One of these water molecules per sulfonate group was a hydronium ion, $H_3O^+$. Systems were built using the Enhanced Monte Carlo (EMC) code[26] at initial densities of 0.8-1.0 g/cm$^3$. Simulations used the DREIDING force field for the polymer, water, and hydronium.[8,44] All systems were equilibrated with classical MD simulations using the LAMMPS[55] package with a timestep of 1 fs and interactions between the Pt atoms and other atoms in the system were described using a classical force field[6].

The bulk systems consisted of just the hydrated Nafion. The composite systems included a platinum slab with 4 atomic layers that was built using the same procedure as for the DFT calculations. Both the bulk and composite systems were subjected to an annealing protocol involving successive cycles of MD simulations at high temperatures and pressures in order to overcome large energy barriers. The protocol used here is similar to that used previously to anneal other hydrated, glassy, ionic polymers, and has been shown to produce systems with reasonable densities as compared with experiment.[1,12,46] We obtained good agreement of the resulting systems with previous simulations of Nafion with the DREIDING force field and the SCP/Fw water model[56] and the densities of our Pt-Nafion-water training set (Figure S1). Further details of the simulations and the annealing protocols can be found in SI Section V.

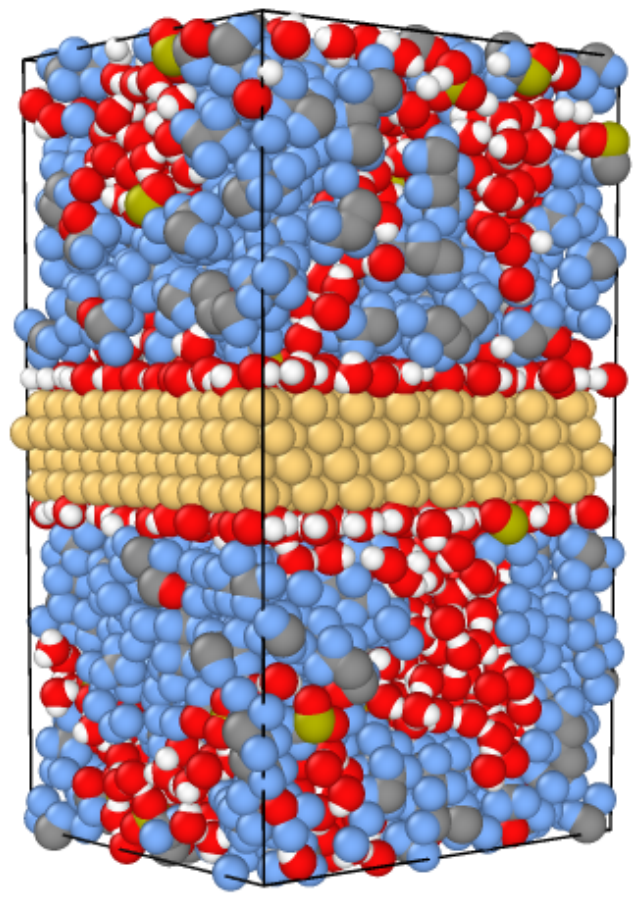

Figure 2. Snapshot from the end of the classical MD annealing runs of the composite system of platinum, Nafion, water, and hydronium studied here, for $\lambda = 15$. Platinum, hydrogen, oxygen, carbon, sulfur, and fluorine atoms are represented as light brown, white, red, gray, gold, and blue spheres, respectively. The structure was rendered using Ovito.[53]

A snapshot of the final configuration after annealing of one of the composite systems is shown in Figure 2. There is a layer of water and some sulfonate groups at each of the Pt surfaces, in qualitative agreement with previous simulations of Pt-Nafion-water interfaces.[29] Following the equilibration procedures, the bulk and composite systems were run for 1 ns using the trained MACE MLIP models at 300 K using NVT dynamics through a Nose-Hoover thermostat. When computing mean-squared displacements (MSDs) for our MACE MLIP trajectories to study hydrogen atom transport, we did not include the first 200 ps of the trajectories to account for enhanced particle movement due to the velocity initialization and switch from the classical MD to the MACE potential. We also attempted to run tests using two foundational models (MATPES-r2SCAN and MPA-0) and found that we had insufficient memory to run them on the higher hydration cells Nafion-water systems, as well as Pt–Nafion-water surface systems, on a single GPU. While multi-GPU implementations of MACE exist, they are in beta testing, so we did not continue this investigation any further.

## III. RESULTS AND DISCUSSION

### A. Structure and transport

In assessing the quality of any MLIP, it is important to compute a range of physical properties to evaluate the accuracy of the model beyond force and energy predictions. In what follows, we compute structural and transport properties using our MLIP and compare them to our DFT data and literature values.

Figures S5 and S6 summarize the bond distance histograms for several atom-type pairs for both the AIMD and our MACE MLIP MD trajectories, respectively, and confirm good model performance. The distance at which the bond length histograms peak for all atom type pairs are nearly identical between our MLIP and DFT. The O-S, C-C, C-F histogram peaks are shifted to slightly longer distances for our model's trajectories, but still differ with the corresponding AIMD peaks by less than 0.01 Å. The histogram widths produced by our model are also in good agreement with the AIMD data, except for O-H and Pt-Pt interactions, which have somewhat larger peak widths. Similar conclusions hold for our model's bond angle predictions, as shown in Figures S8 and S9. To further understand the differences between AIMD and MACE, we carried out simulations using the MACE model beginning from the starting frame of the AIMD simulations and running for 2 ps each as in AIMD. The results in Figure S7 and S10 show that our model closely replicates that AIMD results when run on the same cell for the same amount of time as AIMD.

The Pt-O interaction distance histogram from our model is substantially different than the ones from either the AIMD or classical MD trajectories. Figure S5 shows that the AIMD Pt-O histogram is rather flat, lacking a sharp peak over a Pt-O interaction distance of 1.9-2.7 Å, while the classical MD histogram exhibits a broad peak spanning 2.4-2.7 Å (Figure 3 left). In contrast, our model predicts a much sharper Pt-O bond distance histogram peak at approximately 2.2 Å with a much smaller tail extending out to 2.7 Å (Figure 3 right). The nearly flat Pt-O histogram for the AIMD data arises from the lack of large water displacements from their initial positions during the relatively short 500 fs trajectories used to generate the training data. The broad Pt-O classical MD histogram peak arises from the force field used.[6] The shift in our model's Pt-O bond length histogram to a lower distance with a sharper peak as compared to classical MD shows that it has learned Pt-water interactions and that water migration towards the Pt surface can occur on the 1 ns timescale of our MLIP trajectories.

Next, we computed radial distribution functions (RDFs) using MDAnalysis[21] on output trajectories from the 1 ns MLIP trajectory to understand local vs long-range atomic environments at the different hydration levels. Each RDF quantifies how the radial density of a particular atom type varies with distance from reference atoms of another atom type. With the exception of the sulfur-sulfur (S-S) RDF, we found that the RDFs were not substantially impacted by hydration level, aside from the peak intensities. Figure 4 shows the RDFs of sulfur-sulfur (S-S) and sulfonate oxygen - water or hydronium oxygen ($O_S$-$O_{WH}$) pairs in the bulk (Nafion-water) and composite (Pt-Nafion-water) systems. Examining the S-S RDFs show the first peak around 6 Å. In classical molecular dynamics simulations the first peak is found to be at distances from 4.2 Å[13,50,51] up to 6 Å[15], depending on the calculation details. The trend of slightly right-shifting peak location with increasing hydration is present in both classical MD and with our MLIP. Comparing to the MLIP developed by Jinnouchi, et al.[28], we find approximate agreement with our peak locations and the trend of slightly right-shifting the peaks with increasing hydration, as well as agreement with first-principles (FPMD) based calculations[11]. The relatively constant distribution following the first S-S peak is also exhibited in classical MD, other MLIP work, and FPMD.

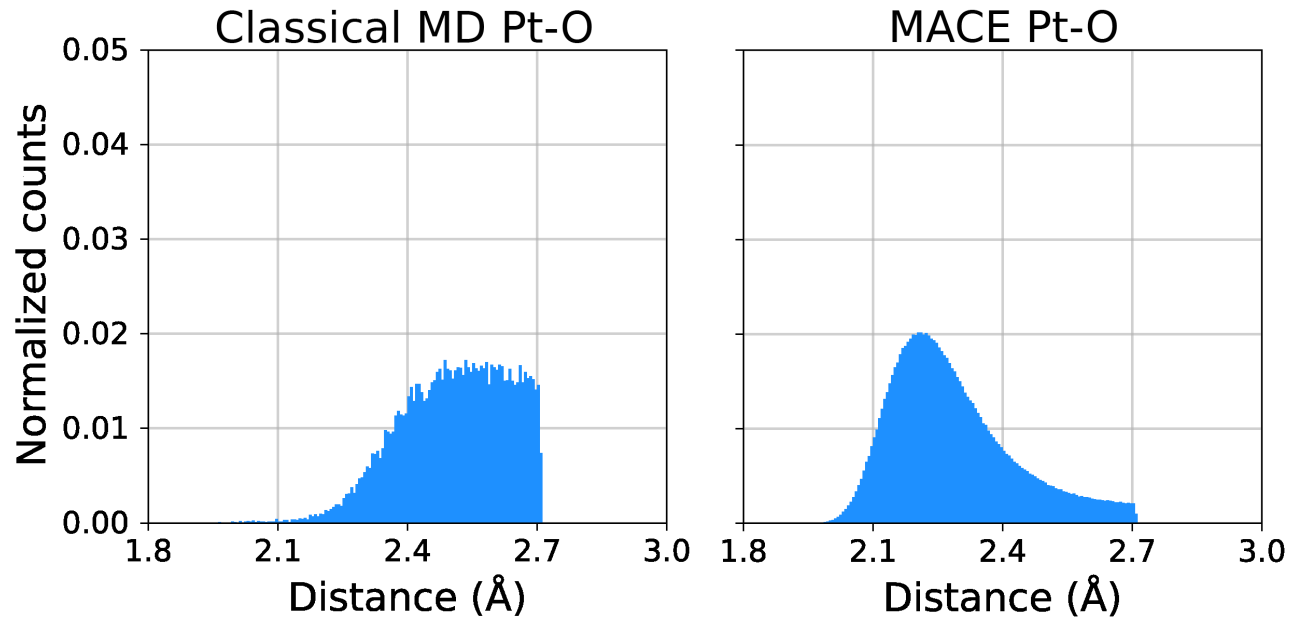

Figure 3. Pt-O bond distances from last 300 ps of classical MD equilibration compared to MACE.

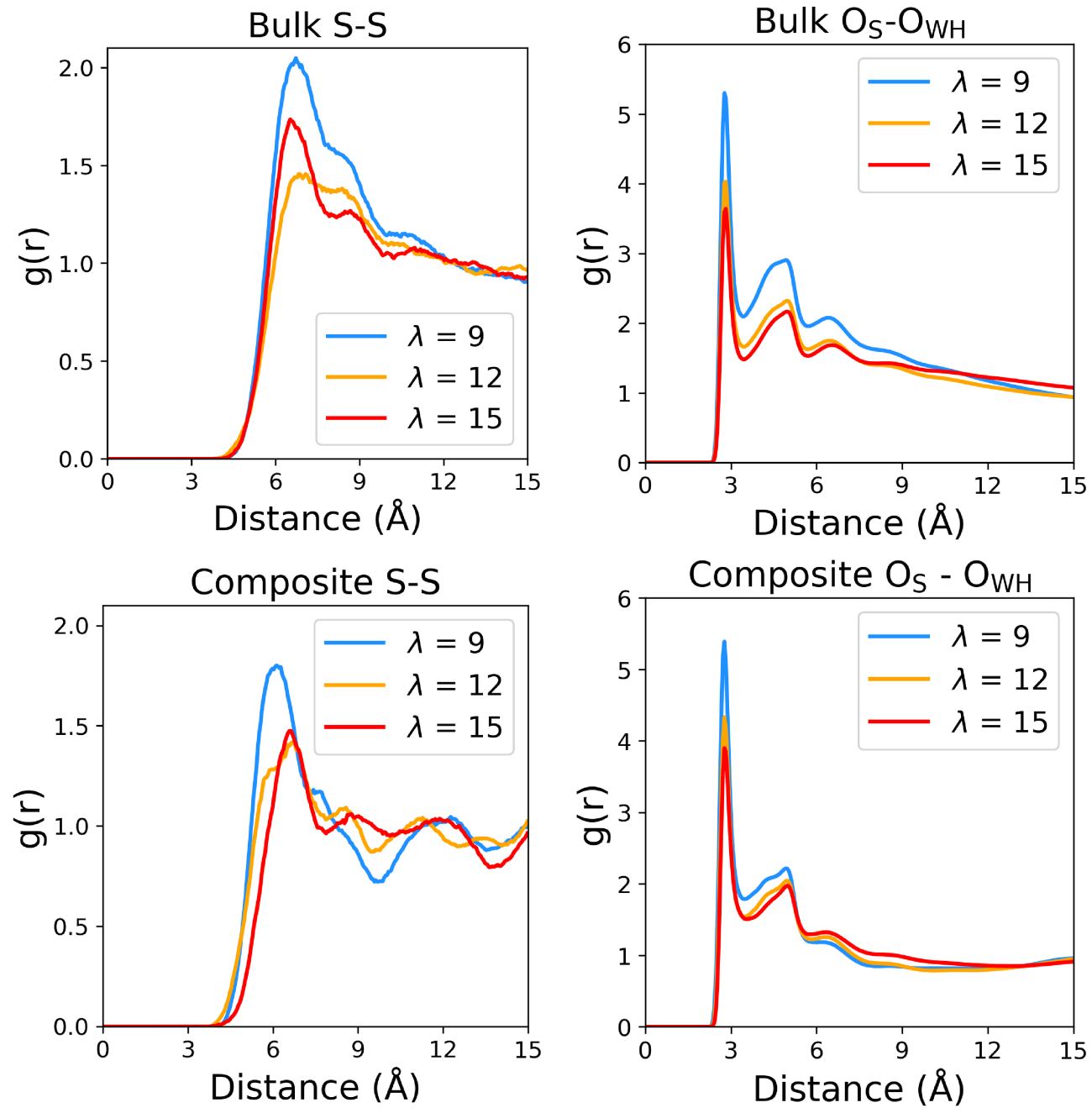

Figure 4. Radial distribution functions (RDFs) for S-S (left) and $O_S$-$O_{WH}$ (right) atom-type pairs. The bulk Nafion-water systems are on top and composite Pt-Nafion-water systems are on bottom, wtih different hydration levels $\lambda$ shows as different line colors. All RDFs were calculated using structures predicted with our MLIP after a 1 ns trajectory.

The first and second peak locations, minima between peaks, and the trend with increasing hydration in the Nafion-water $O_S$-$O_{WH}$ RDFs agree well with classical MD simulations[13,61]. The first peak in the Nafion-water and Pt-Nafion-water $O_S$-$O_{WH}$ RDFs are also similar, but differences arise in the second and third peaks, with the Nafion-water system exhibiting longer-ranged ordering. Additional RDFs seen in S4 are consistent with values seen in literature from classical MD simulations[13,51,60,61] as well, indicating that the trained model is capable of accurately determining and representing system structure. Other RDFs in Figure S4 also show the coordination peaks at larger distance being less pronounced in the composite systems. We hypothesize that the composite systems do not reach the same morphology as the bulk systems due to their small size, which leads to somewhat less order at longer distances.

Since our trained MLIP exhibits good accuracy for structural property prediction, we next evaluated its performance for predicting transport via the calculation of mean-squared displacements (MSDs) for hydrogen atoms using MDAnalysis[21]. Figure S11 shows that all of these systems are still in the sub-diffusive regime after the 1 ns trajectory due to the constrained nature of the water dynamics in the polymer. This is expected as classical MD diffusion constants are frequently calculated from MSDs using trajectories at least dozens of ns long[16,46]. Despite this, Table S6 shows that the hydrogen atom MSDs for the Nafion-water bulk systems were nearly 3 times higher than the MSDs for the Pt-Nafion-water composition systems, suggesting that the surface reduces proton mobility. Additionally, we find that hydrogen MSDs increased with hydration, consistent with typical classical MD diffusion constant predictions.

Two types of MSD were calculated: 1) the MSD of all the hydrogen atoms in the system, and 2) the MSD for protons calculated using the method described in SI Section VII. The second approach was adopted to understand the proton displacement separately from that of the water, and takes Grotthuss hopping seen into account. Plots of the MSDs for these two approaches are shown in Figures S11-S13. The all-hydrogen log-scaled plot is shown in Figure 5. Both all-hydrogen and proton MSDs show a clear trend in the bulk systems, with increasing hydration resulting in increased displacement, as expected. The difference between hydration levels is greater for the proton MSDs than the all-hydrogen MSDs. The composite systems have lower displacements than the bulk systems for both MSDs, but the displacements also increase with increasing hydration except for the outlier $\lambda = 9$ proton MSD discussed below. For the bulk systems the proton displacements are larger than the all-hydrogen displacements at any given time. This is evidence that there are some Grotthuss hopping events occurring in these systems that lead to the protons moving faster than the water. Visual inspection of the simulation trajectories confirm that there are Grotthuss-hopping type events. The $\lambda = 9$ composite system shows a dramatic change in the proton MSD around 400 ps, which is attributed to the hydronium interacting with the surface, displaying the necessity for longer timescales to adequately describe the dynamics of the system.

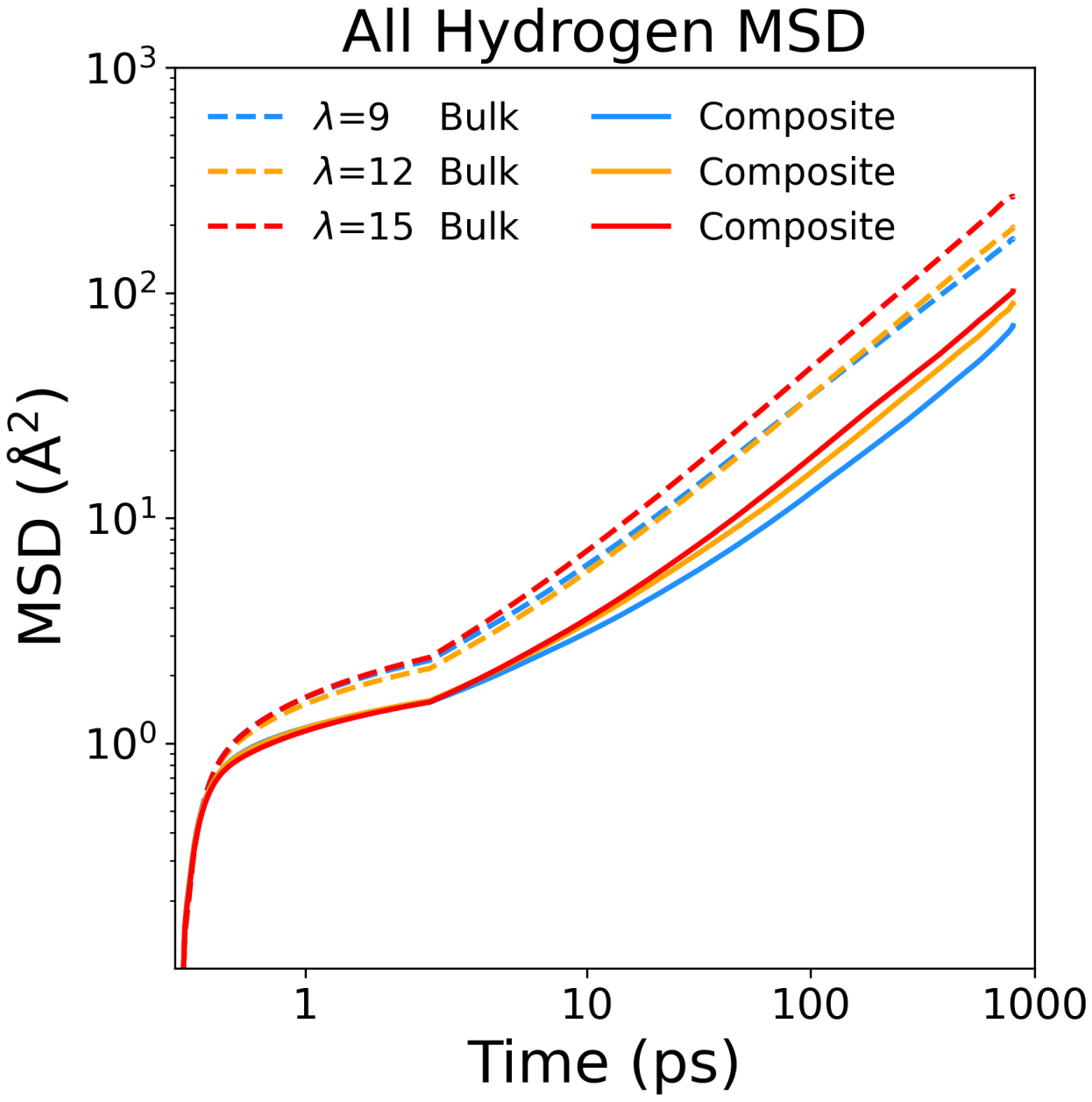

Figure 5. MSD computed including all hydrogen atoms in the systems. This includes hydrogen atoms in the water molecules, free protons, and hydronium ions. Solid (dotted) lines are for the composite (bulk) systems, and different hydration $\lambda$ are specified by different line colors.

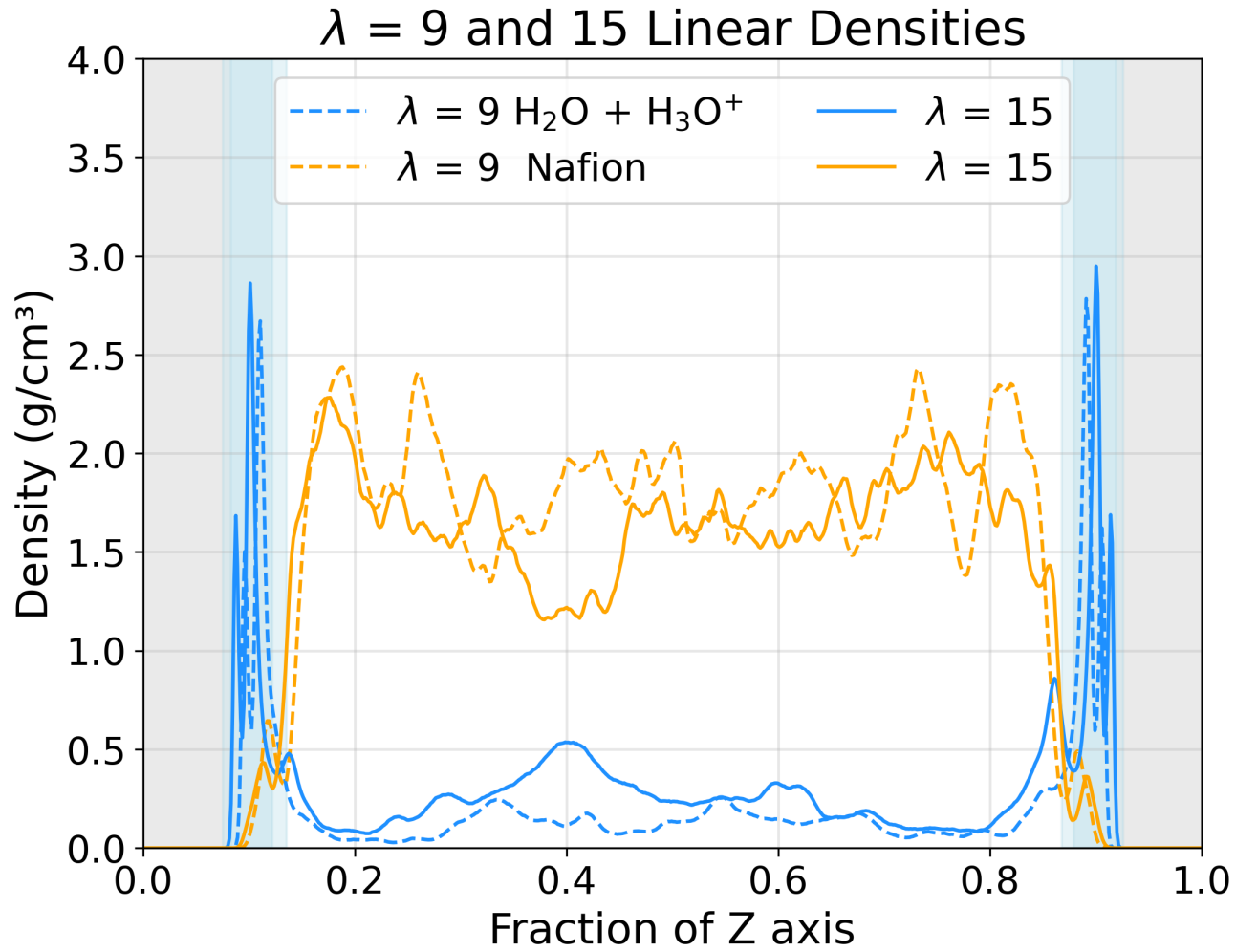

Figure 6. Z-profiles of the planar-averaged mass density of Nafion and water and hydronium for $\lambda = 9$ and $\lambda = 15$ Pt-Nafion-water composite systems. The system models are fully periodic, with the distance between periodic surfaces approximately 60-70 Å. The portion of the z-axis consisting of the Pt(111) surface in the structures is represented by the gray shaded regions. Here, the z axis is translated so that the middle of the Pt(111) surface has a coordinate of 0, and is normalized to compare the systems with different z-axis lengths. The surface region dominated by water is shaded blue and was defined to go from the Pt(111) surface until the first trough in the water+hydronium density profile after the largest density peak.

To further investigate why the water and protons move more slowly in the Pt-Nafion-water systems than in the Nafion-water systems, we calculated planar-averaged densities of Nafion and water and hydronium across the z-axis of the Pt-Nafion-water systems. Figure 6 summarizes results for $\lambda = 9$ and $\lambda = 15$, while Figures S14 and S15 show complete results for $\lambda = 9$, 12, and 15. We find that the MLIP predicts an interfacial region between the Pt and Nafion dominated by water and hydronium. Additionally, increasing hydration from 9 to 15 does not contribute further water to the interface, but instead increases water content within the bulk-like regime of the systems where Nafion dominates the density profiles. The preferential water interaction with the Pt reduces the effective hydration level within the Nafion, and thus lowers the expected proton MSD over a trajectory.

Interestingly, similar behavior has also been observed by Kang, et al.,[29] in a classical MD study using a DREIDING force field for Nafion, an embedded atom model for Pt, the FC3 force field for water[40], and force field parameters from Brunello, et al. for the interactions of Pt with all other atoms[6] They found that increasing system water content from a hydration of 3 to 10 greatly increased water present in this interfacial region, but that this region saturates and further water hydrates the bulk-like interior of the Nafion. We see similar behavior, with the amount of water in the interfacial region being approximately constant as $\lambda$ increases from 9 to 15 (Figure S12). Kang, et al. find that the interfacial region contains both water and Nafion, whereas we find that the first layer next to the Pt is mostly water. This difference could either be due to our MLIP learning that Pt-water interactions are lower energy or that the Nafion chains did not have time to fully equilibrate during the 1 ns trajectories used here, or both, as discussed in more detail above. We conclude that our model likely describes proton transport well within the limitations of current MLIP architectures in simulating long MD trajectories.

### B. Proton-transfer and polymer dissociation reaction pathways

Our MLIP is reactive and predicts proton motion via a combination of vehicular transport and Grotthuss hopping. We thus created a set of 12 proton-transfer and polymer dissociation reaction pathways that could be benchmarked against r2SCAN DFT energies to benchmark its performance for stretched bond configurations. We also compared to energies from the medium MACE-MATPES-r2SCAN-0 model, medium MACE-MPA-0 foundational model[4], and Perdew-Burke-Ernzerhof (PBE) functional[47]. The reaction set includes 1) Grotthuss hopping via $H_5O_2$ transfer or $H_3O$ creation, 2) $SO_3$ group deprotonation via $H_5O_2$, $H_3O$, $H_2O$, or $H^*$ species, 3) polymer dissociation with either $SO_3$, $CF_3$, or an F atom as the leaving group, 4) water desorption from the Pt(111) surface or water dissociation into $H^*$ and $OH^*$, and 5) $O_2$ adsorption and protonation to $OOH^*$ by a nearby $HSO_3$ group, which is a first possible step in the oxygen reduction reaction (ORR). This set spans reactions well within the training set (e.g. $SO_3$ group deprotonation to form $H_3O$) and outside the training set (e.g. $O_2^*$ protonation to $OOH^*$ or F atom dissociation from Nafion).

These reaction pathways were generated using the Atomic Simulation Environment[39] within a $\lambda$=12 Pt-Nafion-water AIMD structure with no strain applied. Only the atoms involved in each reactant/product state were relaxed using r2SCAN to isolate each reaction's impact on system energy. Intermediate image energies were calculated without ion optimization because exact transition states are not needed to evaluate stretched bond configurations. Additional details on reaction pathway generation are in SI Section IX.

Figure 7 shows results for three representative reaction pathways and a summary of the root mean squared error (RMSE) averaged across all 12 reactions for each of the 10 reaction coordinate images, providing an estimate for general model accuracy. Figures S16 and S17 show the performance of the model for all 12 reactions individually and the RMSE averaged across all images within each reaction. When compared to the r2SCAN energies, our MLIP performs well (overall RMSE = 0.16 eV), exhibiting somewhat better accuracy as the medium MACE-MPA-0 model predictions compared against PBE energies (overall RMSE = 0.21 eV) and the MACE-MATPES-r2SCAN-0 model predictions compared against r2SCAN energies (overall RMSE = 0.30 eV). As expected, our model is most accurate for the Grotthuss-type reactions well-represented in the training data, such as for Zundel species migration (Figure 7A) or $SO_3$ group deprotonation by $H_3O$ (Figure 7B), and predicts these image energies with an overall RMSE of less than 0.03 eV across the pathways. Interestingly, our model exhibits good accuracy for the surface reactions of $O_2^*$ protonation to $OOH^*$ along with $H_2O^*$ dissociation into $H^*$ and $OH^*$ (Figure 7C) despite having no training data included on these adsorbed species, suggesting it has learned additional information about O and H bonding beyond the targeted training set. It also described the $SO_3$ and $CF_3$ dissociation reactions away from the Nafion chain well (Figure S16), consistent with the lack of polymer dissociation observed over the MLIP's 1 ns trajectory. The model was least accurate for intermediate and product images along the pathways that contain a free H, OH, or F (Figure S16). This is unsurprising given the lack of these species in the training data.

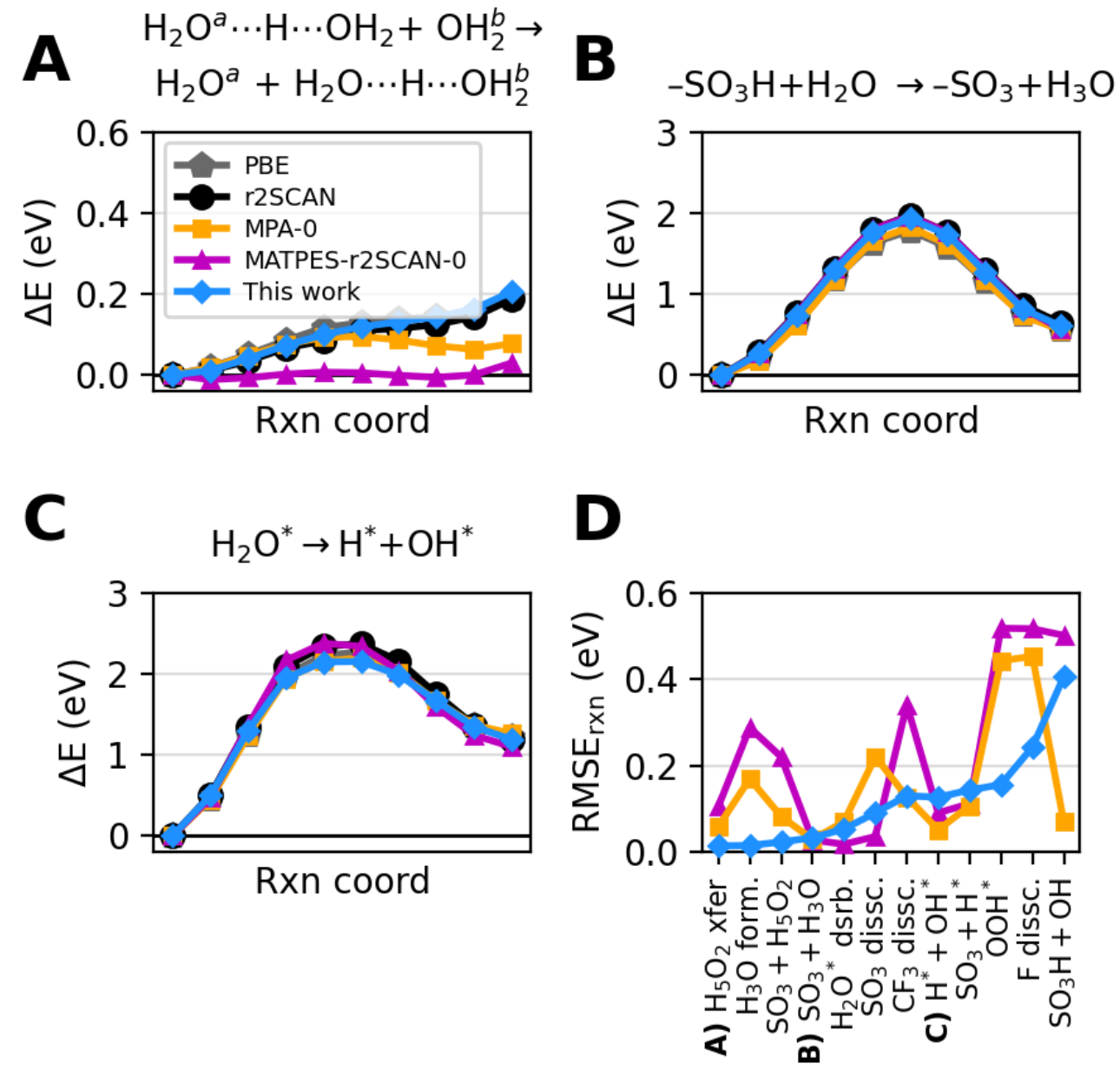

Figure 7. A-C) Representative reaction pathway energies predicted by the MLIP model trained in this work compared to r2SCAN energies, the MACE-MATPES-r2SCAN-0 medium model, MACE-MPA-0 medium model, and PBE calculations. We note that the starting reactant state (left-most point) for each reaction had 0 error because reaction pathway energies are shifted such that each of these values are always 0 eV. The net reaction energy in subfigure A is not 0 because the non-participating atoms were frozen and the participating atoms are in somewhat different surrounding environments, despite having the same chemical formulas on each side of the reaction. D) Reaction pathway energy root mean squared error (RMSE) averaged over all 10 images in each reaction pathway for the MLIP model trained in this work compared to the MACE-MATPES-r2SCAN-0 medium model, MACE-MPA-0 medium model, and r2SCAN DFT. The starting state for each reaction was excluded from the RMSE calculations shown in this plot because its energy was always shifted to be 0 eV. See the main text for a more detailed description of the reaction set and Figure S16 for an alternate version of this figure with more detailed labels. The reactions for subplots A-C are each marked here in the corresponding shortened x-axis label.

Figures 7D, S16, and S17 show that the RMSEs of all models is substantially impacted by inaccuracies for 3 reactions, namely $O_2^*$ protonation to $OOH^*$, F atom dissociation from $CF_3$, and $SO_3$ protonation by $H_2O$ (leaving

a free OH group). Removing these 3 reactions nearly halves the overall models' RMSEs to just 0.18 eV for MACE-MATPES-r2SCAN-0, 0.12 eV for MACE-MPA-0, and 0.09 eV for our model. Lastly, the impact of DFT functional on training can be seen in Figure S18, which shows that for these 12 reactions r2SCAN tends to predict approximately 0.05-0.15 eV higher intermediate image energies. Such a disagreement for converged barrier height predictions could impact rate constant predictions by approximately a factor of 7-330x at 300 K, and so is important to consider in future MLIP training efforts.

Overall, our MLIP model can perform well for interpolative tasks with a degree of extrapolative accuracy, but its applicability to structures well outside the training set is somewhat limited. Although challenging to accomplish for large configurational spaces, the model could likely be most improved by additional sampling of highly stretched bonds and higher energy states in the potential energy surface using techniques such as contour mapping[58]. Additional error quantification of model performance via protocols that automatically generate many reaction pathways of interest would also be useful.

## IV. CONCLUSIONS

In this work, we have developed a MLIP to simultaneously treat polymer structure, transport, and chemical reactions in a system of hydrated Nafion on platinum using the MACE approach. The structure of and transport through Nafion are generally well described, with trends that match existing results in the literature. Our model finds a significantly shorter Pt-O bond length than the classical potential of Brunello et al.,[6] likely due to stronger water coordination with the platinum surface. The motion of protons is found to be slower in our composite system with a platinum surface than in bulk Nafion due to the high density of water near the platinum, as expected. Additionally, we find that our MLIP-predicted proton transport contains both vehicular transport and Grotthuss hopping. Further investigation with calculations over longer times are needed to calculate diffusivities and ascertain the accuracy of model for these predictions. For reaction pathways along the Pt surface, we find the MLIP performs well for reactions for which the products are contained in the training set, with significantly less accuracy when they are not. The MACE MPA-0 and MACE-MATPES-r2SCAN-0 foundational models show moderately lower accuracy for the reactions than our MLIP, although MPA-0's performance for this task is impressive given its training set. In comparison to these foundational models, our MLIP is a lighter weight model, allowing systems to be run faster with less memory. Active learning does not yield any improvement for our model.

There are several avenues for improvement for the current model. The first general area involves the expansion of the training set. For the description of reactions, the addition of different reactants and products to our training set, as well as transition states between them, is expected to improve accuracy to desired levels. Recent approaches such as the random exploration via imaginary chemicals optimization (REICO) sampling strategy[59], are appealing because they could lead to improvements in predictions of both thermodynamic ground states and kinetics without explicitly including the exact states in the training set. Regardless of the particular approach, more innovative methods are needed to explore the large phase space of electronic interactions characteristic of heterogeneous systems, as indicated by the lack of improvement of our model with traditional active learning schemes. Additionally, the relatively robust results of the MP-0 foundational model for different reactions explored here indicate that fine-tuning this foundational model may also be a promising approach to describing this and related systems. All system properties predicted are likely to benefit from using a higher-level functional for the training data, e.g. the random phase approximation (RPA)[18,38]. Many improvements in the training and prediction will be enabled by the continued expansion of computing power, as well as the continued development of more efficient and scalable MLIPs.

The performance of many chemical devices and processes depend sensitively on structure, transport and reactivity, and their interplay, so predicting all three properties in a single model could significantly impact the development of many technologies. The model developed here shows the promise of state-of-the-art MLIP architectures and computing resources to simulate the complexity of these devices. In comparison to DFT, the larger simulation cells and longer simulation times possible with MLIPs allow for more accurate representations of complex systems and their dynamics, as well as comparison to experimental observables. The efficient representation of quantum mechanical atomic interactions enables the accurate treatment of arbitrary interfaces, reaction energetics and processes such as Grotthuss hopping over these large length and time scales. The development of robust models that sample all relevant phenomena for a given process or technology is challenging, but is expected to become more straightforward and even autonomous with the aide of AI. While significant work is needed to achieve the autonomous development of MLIPs that are accurate for any chemical technology of interest, modern MLIP architectures and computing technology have brought this once impossible task within reach.

## V. ACKNOWLEDGMENTS

We acknowledge financial support from the US Department of Energy Office of Critical Minerals and Energy Innovation, Hydrogen and Fuel Cell Technologies Office, under the ElectroCat Consortium, DOE technology manager William Gibbons, and DOE program manager Dimitrios Papageorgopolous. This work was authored by the National Laboratory of the Rockies, operated by the Alliance for Energy Innovation, LLC, for the U.S. Department of Energy (DOE) under Contract No. DE-AC36-08GO28308. The research was performed using computational resources sponsored by the Department of Energy's Office of Critical Minerals and Energy Efficiency and located at the National Laboratory of the Rockies, and at the Center for Computational Innovations at Rensselaer Polytechnic Institute. This work was performed, in part, at the Center for Integrated Nanotechnologies, an Office of Science User Facility operated for the U.S. Department of Energy (DOE) Office of Science. Sandia National Laboratories is a multimission laboratory managed and operated by National Technology and Engineering Solutions of Sandia, LLC, a wholly owned subsidiary of Honeywell International, Inc., for the U.S. DOE's National Nuclear Security Administration under contract DE-NA-0003525. The views expressed in the article do not necessarily represent the views of the U.S. DOE or the United States Government. We would like to thank the National Nuclear Security Administration IMPACT (NNSA-IMPACT) program for providing the opportunity to Samuel Brown to work on part of the work presented.

# Supporting Information: Development of machine-learned interatomic potentials to predict structure, transport, and reactivity in platinum-based fuel cells

Kamron Fazel,[1,a)] Samuel Brown,[2,a)] Jacob Clary,[3] Pritom Bose,[4] Nima Karimitari,[5] Amalie L. Frischknecht,[6] Ravishankar Sundararaman,[1,b)] and Derek Vigil-Fowler[3,c)]

[1)] *Materials Science & Engineering, Rensselaer Polytechnic Institute, Troy, NY 12180, USA*

[2)] *Department of Chemistry and Biochemistry, New Mexico State University, Las Cruces, NM 88003, USA*

[3)] *Materials, Chemical, and Computational Science Directorate, National Renewable Energy Laboratory, Golden, CO 80401, USA*

[4)] *Mechanical, Aerospace & Nuclear Engineering, Rensselaer Polytechnic Institute, Troy, NY 12180, USA*

[5)] *Department of Chemistry and Biochemistry, University of South Carolina, Columbia, SC 29208, USA*

[6)] *Center for Integrated Nanotechnologies, Sandia National Laboratories, Albuquerque, New Mexico 87185, USA*

[a)] These authors contributed equally to this work.

[b)] Electronic mail: sundar@rpi.edu

[c)] Electronic mail: derek.vigil-fowler@nrel.gov

The Supporting Information contains additional summaries of the system properties of the AIMD systems used in initial model training, details on the active learning approach, additional DFT computational details, additional summary histograms of the AIMD and our MLIP bond lengths and angles for the Pt-Nafion-water composite system, more detail on our proton tracking methodology, additional planar-averaged density profiles of the composite system, and additional reaction pathway results.

## I. MACE TRAINING DATA

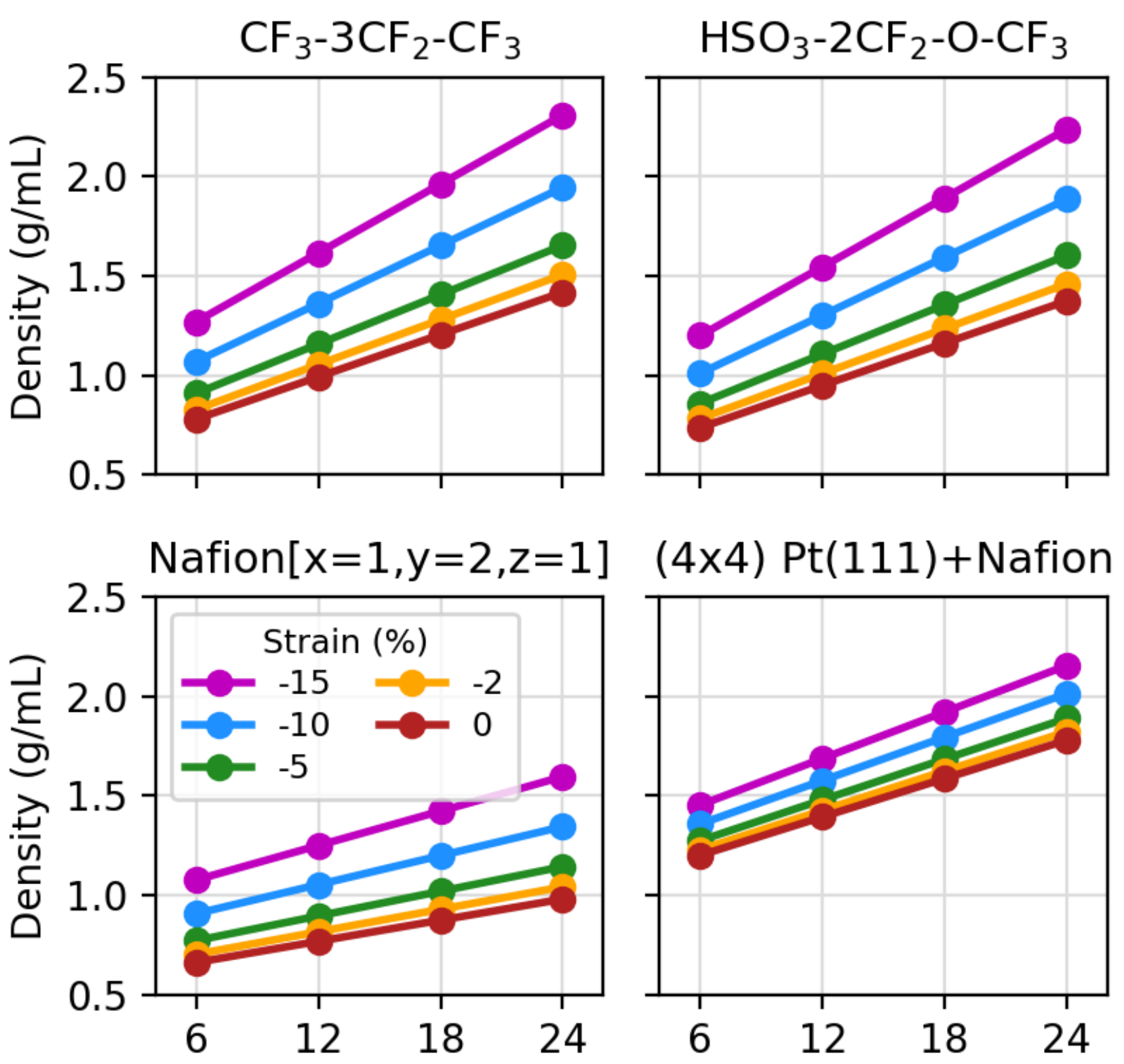

Figure S1: Densities of the Nafion+water region of all systems containing Nafion in the training set. For systems without Pt, the volume used is simply the total volume of the cell, $V_{\text{total}}$. For systems with Pt, the volume occupied by Nafion+water is defined as: $\mathrm{V} = \mathrm{V}_{\text{total}} - \mathrm{A}_{\text{xy}}\mathrm{H}_{\text{slab}}$, where $\mathrm{H}_{\text{slab}}$ is the center to center distance between top and bottom Pt rows plus double the covalent radius of a Pt atom.

$$-\left[\left[CF_2-CF_2\right]_x-CF\left(-\left[O-CF_2-CF(CF_3)\right]_z-O-CF_2-CF_2-SO_3H\right)-CF_2\right]_y-$$

Figure S2: Nafion chemical structure used to generate AIMD water-Nafion and Pt-water-Nafion training data with x = 1, y = 2, and z = 1. An F atom was added to the terminating $CF_2$ groups along the backbone. Each AIMD-simulated system included 1 Nafion chain with a hydration level of 12 $H_2O$ per $SO_3$ tail group. One $SO_3$ tail group was initialized deprotonated, with one water molecule correspondingly replaced by hydronium.

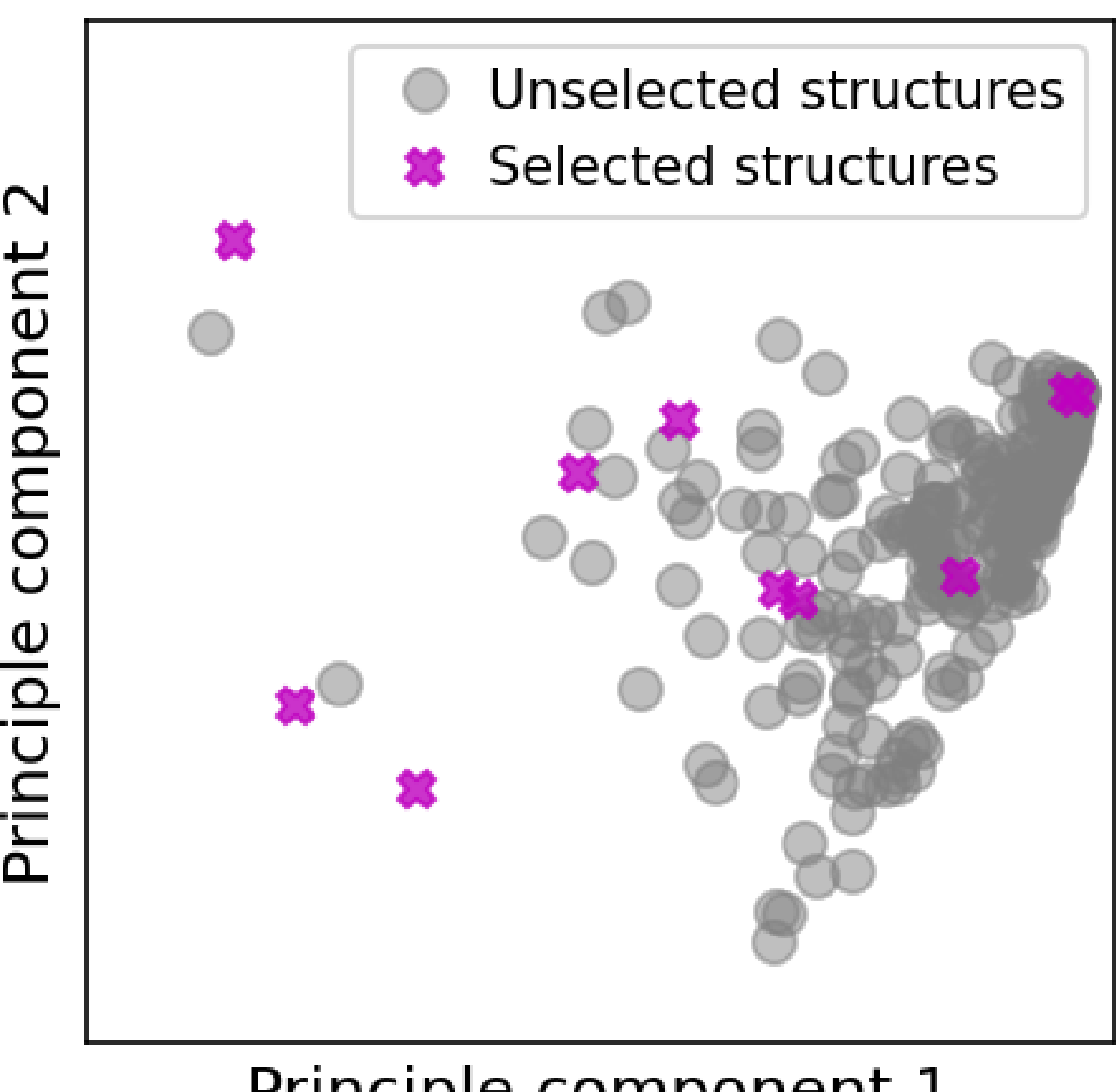

Figure S3: Principle component analysis plot for different Nafion[x=1,y=2,z=1] rotations over a (4×4) Pt(111) surface. The gray circles denote rotations there were not selected for AIMD runs while the magenta Xs denote the 10 rotations selected for AIMD runs with different hydration levels and strains.

## II. MACE TRAINING

The following parameters were used to train the MACE models, with the only difference between the committee of models being the seed.

```
mace_run_train \
    --name="mace-r2-step10-300k" \
    --train_file="train.xyz" \
    --energy_key energy \
    --forces_key forces \
    --valid_fraction=0.03 \
    --config_type_weights='{"Default":1.0}' \
    --E0s="{1:-13.600891103331842, \
            6:-147.67494801444815, \
            8:-429.8144574591154, \
            9:-647.0870118967323, \
            16:-275.5108875572458, \
            78:-3295.2788163081627}" \
    --model="MACE" \
    --hidden_irreps='128x0e + 128x1o' \
    --r_max=5.0 \
    --batch_size=8 \
    --max_num_epochs=500 \
    --ema \
    --ema_decay=0.99 \
    --amsgrad \
    --distributed \
    --restart_latest \
    --save_all_checkpoints \
    --device=cuda \
    --seed=1337
```

## III. MACE ACTIVE LEARNING

Equation S1 was used to determine the relative force deviation for each atom in the active learning process.

$$\sigma_{\text{relative}}(i) = \frac{\sigma(|F_i|)}{\langle |F_i| \rangle + 0.2} \tag{1}$$

where $\sigma_{\text{relative}}(i)$ is the $i$-th particle's relative deviation, $\sigma(|F_i|)$ is the $i$-th particle's force magnitude standard deviation across the committee, and $\langle |F_i| \rangle$ is the $i$-th particle's average force magnitude across the committee. The additive 0.2 is to avoid division by a near zero average force magnitude.

During each active learning iteration, one of the three models was used to produce 24 separate 1 ps NVT trajectories, where each of the trajectories was initialized using the last snapshot of a corresponding training AIMD trajectory as the starting point. The 24 initial configurations used were water at both 0 and -5% strain, Nafion-water systems at -10% linear strain and both $\lambda$=12 and 18, and Pt-Nafion-water systems with all 10 polymer rotations at -10% linear strain and both $\lambda$=12 and 18. Following generation of each trajectory, the other two models were used to predict the forces and energies for all frames in each same trajectory. The frames with a maximum relative force magnitude deviation, calculated using equation S1, between 0.0600 and 0.0625 across all 3 models were then flagged for DFT validation. The DFT calculations on these frames used the same settings as described above and were used to generate new forces and energies added to the training set for the next iteration of models.

Table S1: Initial iteration active learning flagged frames, average per-frame maximum relative force magnitude deviation, average per-frame relative force magnitude deviation, and average per-frame energy deviation per atom (meV/atom). Relative force deviations were calculated using equation S1. All Nafion and Pt-Nafion systems were run with -10% linear strain, while water was run with 0 and -5% strain. All systems were run at 300 K using Langevin dynamics.

| **System** | **Rotation Index** | **Hydration** | **Flagged Frames** | $\langle \sigma_{\mathrm{rel,max}}(\lvert F \rvert) \rangle$ | $\langle \sigma_{\mathrm{rel}}(\lvert F \rvert) \rangle$ | $\langle \sigma(\lvert E \rvert) \rangle$ |
|---|---|---|---|---|---|---|
| Water, 0% strain | – | – | 0 | 0.0093 | 0.0021 | 0.211 |
| Water, -5% strain | – | – | 0 | 0.0099 | 0.0022 | 0.277 |
| Nafion | – | 12 | 0 | 0.0220 | 0.0045 | 1.775 |
| Nafion | – | 18 | 1 | 0.0200 | 0.0039 | 1.420 |
| Pt+Nafion | 1 | 12 | 7 | 0.0327 | 0.0065 | 0.187 |
| Pt+Nafion | 1 | 18 | 10 | 0.0345 | 0.0067 | 0.117 |
| Pt+Nafion | 2 | 12 | 11 | 0.0327 | 0.0065 | 0.173 |
| Pt+Nafion | 2 | 18 | 16 | 0.0366 | 0.0068 | 0.266 |
| Pt+Nafion | 3 | 12 | 108 | 0.0584 | 0.0091 | 0.340 |
| Pt+Nafion | 3 | 18 | 97 | 0.0790 | 0.0097 | 0.291 |
| Pt+Nafion | 4 | 12 | 107 | 0.0661 | 0.0094 | 0.178 |
| Pt+Nafion | 4 | 18 | 24 | 0.0409 | 0.0071 | 0.127 |
| Pt+Nafion | 5 | 12 | 2 | 0.0319 | 0.0067 | 0.243 |
| Pt+Nafion | 5 | 18 | 16 | 0.0367 | 0.0067 | 0.150 |
| Pt+Nafion | 6 | 12 | 30 | 0.0407 | 0.0074 | 0.181 |
| Pt+Nafion | 6 | 18 | 28 | 0.0428 | 0.0078 | 0.133 |
| Pt+Nafion | 7 | 12 | 14 | 0.0345 | 0.0066 | 0.198 |
| Pt+Nafion | 7 | 18 | 13 | 0.0369 | 0.0069 | 0.108 |
| Pt+Nafion | 8 | 12 | 16 | 0.0388 | 0.0074 | 0.178 |
| Pt+Nafion | 8 | 18 | 20 | 0.0370 | 0.0070 | 0.115 |
| Pt+Nafion | 9 | 12 | 8 | 0.0336 | 0.0068 | 0.166 |
| Pt+Nafion | 9 | 18 | 9 | 0.0364 | 0.0072 | 0.161 |
| Pt+Nafion | 10 | 12 | 66 | 0.0477 | 0.0080 | 0.179 |
| Pt+Nafion | 10 | 18 | 70 | 0.0501 | 0.0079 | 0.184 |
| **Average** | | | | **0.0379** | **0.0067** | **0.307** |

Table S2: Iteration 1 active learning flagged frames, average per-frame maximum relative force magnitude deviation, average per-frame relative force magnitude deviation, and average per-frame energy deviation per atom (meV/atom). Relative force deviations were calculated using equation S1. All Nafion and Pt-Nafion systems were run with -10% linear strain, while water was run with 0 and -5% strain. All systems were run at 300 K using Langevin dynamics.

| **System** | **Rotation Index** | **Hydration** | **Flagged Frames** | $\langle \sigma_{\mathrm{rel,max}}(\lvert F \rvert) \rangle$ | $\langle \sigma_{\mathrm{rel}}(\lvert F \rvert) \rangle$ | $\langle \sigma(\lvert E \rvert) \rangle$ |
|---|---|---|---|---|---|---|
| Water, 0% strain | – | – | 0 | 0.0098 | 0.0022 | 0.524 |
| Water, -5% strain | – | – | 0 | 0.0104 | 0.0023 | 0.369 |
| Nafion | – | 12 | 0 | 0.0230 | 0.0048 | 1.398 |
| Nafion | – | 18 | 0 | 0.0208 | 0.0040 | 1.140 |
| Pt+Nafion | 1 | 12 | 9 | 0.0318 | 0.0066 | 0.059 |
| Pt+Nafion | 1 | 18 | 9 | 0.0328 | 0.0065 | 0.079 |
| Pt+Nafion | 2 | 12 | 3 | 0.0307 | 0.0064 | 0.119 |
| Pt+Nafion | 2 | 18 | 8 | 0.0351 | 0.0066 | 0.101 |
| Pt+Nafion | 3 | 12 | 34 | 0.0449 | 0.0077 | 0.144 |
| Pt+Nafion | 3 | 18 | 67 | 0.0712 | 0.0090 | 0.090 |
| Pt+Nafion | 4 | 12 | 38 | 0.0444 | 0.0079 | 0.071 |
| Pt+Nafion | 4 | 18 | 14 | 0.0384 | 0.0072 | 0.060 |
| Pt+Nafion | 5 | 12 | 7 | 0.0313 | 0.0066 | 0.087 |
| Pt+Nafion | 5 | 18 | 8 | 0.0330 | 0.0066 | 0.126 |
| Pt+Nafion | 6 | 12 | 16 | 0.0379 | 0.0071 | 0.049 |
| Pt+Nafion | 6 | 18 | 28 | 0.0404 | 0.0074 | 0.052 |
| Pt+Nafion | 7 | 12 | 4 | 0.0322 | 0.0064 | 0.052 |
| Pt+Nafion | 7 | 18 | 11 | 0.0347 | 0.0067 | 0.097 |
| Pt+Nafion | 8 | 12 | 7 | 0.0362 | 0.0071 | 0.082 |
| Pt+Nafion | 8 | 18 | 7 | 0.0358 | 0.0069 | 0.097 |
| Pt+Nafion | 9 | 12 | 5 | 0.0343 | 0.0069 | 0.073 |
| Pt+Nafion | 9 | 18 | 15 | 0.0348 | 0.0069 | 0.109 |
| Pt+Nafion | 10 | 12 | 42 | 0.0438 | 0.0074 | 0.091 |
| Pt+Nafion | 10 | 18 | 55 | 0.0455 | 0.0074 | 0.083 |
| **Average** | | | | **0.0347** | **0.0064** | **0.215** |

Table S3: Iteration 2 active learning flagged frames, average per-frame maximum relative force magnitude deviation, average per-frame relative force magnitude deviation, and average per-frame energy deviation per atom (meV/atom). Relative force deviations were calculated using equation S1. All Nafion and Pt-Nafion systems were run with -10% linear strain, while water was run with 0 and -5% strain. All systems were run at 300 K using Langevin dynamics.

| **System** | **Rotation Index** | **Hydration** | $\langle \sigma_{\mathrm{rel,max}}(\lvert F \rvert) \rangle$ | $\langle \sigma_{\mathrm{rel}}(\lvert F \rvert) \rangle$ | $\langle \sigma(\lvert E \rvert) \rangle$ |
|---|---|---|---|---|---|
| Water, 0% strain | – | – | 0.0096 | 0.0022 | 1.220 |
| Water, -5% strain | – | – | 0.0102 | 0.0023 | 1.159 |
| Nafion | – | 12 | 0.0217 | 0.0045 | 0.521 |
| Nafion | – | 18 | 0.0199 | 0.0039 | 0.358 |
| Pt+Nafion | 1 | 12 | 0.0320 | 0.0065 | 0.115 |
| Pt+Nafion | 1 | 18 | 0.0354 | 0.0066 | 0.086 |
| Pt+Nafion | 2 | 12 | 0.0315 | 0.0064 | 0.112 |
| Pt+Nafion | 2 | 18 | 0.0340 | 0.0066 | 0.103 |
| Pt+Nafion | 3 | 12 | 0.0430 | 0.0075 | 0.116 |
| Pt+Nafion | 3 | 18 | 0.0567 | 0.0080 | 0.121 |
| Pt+Nafion | 4 | 12 | 0.0454 | 0.0079 | 0.086 |
| Pt+Nafion | 4 | 18 | 0.0386 | 0.0071 | 0.063 |
| Pt+Nafion | 5 | 12 | 0.0314 | 0.0066 | 0.109 |
| Pt+Nafion | 5 | 18 | 0.0319 | 0.0065 | 0.125 |
| Pt+Nafion | 6 | 12 | 0.0364 | 0.0069 | 0.109 |
| Pt+Nafion | 6 | 18 | 0.0398 | 0.0073 | 0.070 |
| Pt+Nafion | 7 | 12 | 0.0320 | 0.0066 | 0.121 |
| Pt+Nafion | 7 | 18 | 0.0370 | 0.0069 | 0.089 |
| Pt+Nafion | 8 | 12 | 0.0355 | 0.0071 | 0.108 |
| Pt+Nafion | 8 | 18 | 0.0343 | 0.0068 | 0.085 |
| Pt+Nafion | 9 | 12 | 0.0348 | 0.0068 | 0.104 |
| Pt+Nafion | 9 | 18 | 0.0363 | 0.0071 | 0.086 |
| Pt+Nafion | 10 | 12 | 0.0407 | 0.0071 | 0.118 |
| Pt+Nafion | 10 | 18 | 0.0434 | 0.0071 | 0.106 |
| **Average** | | | **0.0338** | **0.0063** | **0.220** |

Table S4: Energy (meV/atom) and force (meV/Å) evaluation error of our trained MACE models evaluated on a common test set.

| **Iteration** | $\langle \mathrm{RMSE}(E) \rangle$ | $\langle \mathrm{RMSE}(F) \rangle$ |
|---|---|---|
| Preliminary | 3.50 | 49.17 |
| Iteration 1 | 3.43 | 49.23 |
| Iteration 2 | 3.90 | 49.33 |

Table S5: Ablation test energy (meV/atom) and force (meV/Å) error with varying training dataset compositions. The first row shows the errors for the full dataset, while subsequent rows show the errors when we removed the part of the full dataset indicated in the "Removal from Full Training Dataset" column, e.g. row 4 gives the errors when we removed the training data of the Nafion water systems.

| **Removal from Full Training Dataset** | $\langle \mathrm{RMSE}(E) \rangle$ | $\langle \mathrm{RMSE}(F) \rangle$ |
|---|---|---|
| None | 3.5 | 49.17 |
| Half of Nafion Rotations on Platinum | 3.4 | 63.4 |
| Strained systems | 8.0 | 189.7 |
| Nafion-water | 11.3 | 59.5 |
| Strained systems except -15% | 4.2 | 60.1 |

## IV. ADDITIONAL DFT COMPUTATIONAL DETAILS

The Pt surface was built by expanding the fcc Pt bulk primitive cell into a (4×4) supercell with 4 layers and 64 total atoms. We used a 2×2×1 Γ-centered **k**-grid and 0.2 eV Fermi electronic smearing for all systems containing the (4×4) Pt(111) surface and a 1×1×1 Γ-centered **k**-grid for systems not containing Pt. Each AIMD trajectory in the training data set was generated using the NVT ensemble at 300 K using a Nose-Hoover thermostat with 1 fs timesteps. The wavefunctions were converged to an energy threshold of $10^{-6}$ hartrees at each time step.

## V. DETAILS OF THE CLASSICAL MD SIMULATIONS

Temperature and pressure were controlled with a Nose-Hoover thermostat and barostat with time constants of 0.1 ps and 1 ps, respectively. A short range cutoff of 10 Å was used for intermolecular interactions, while long-range Coulomb interactions were computed using the particle-particle particle-mesh (PPPM) algorithm with an accuracy of $10^{-5}$.

For the Pt-Nafion-water systems, a platinum slab with 4 atomic layers was built using the same procedure as for the DFT calculations, and had lateral dimensions of $L_x = 33.2$ Å and $L_y = 38.3$ Å. The Pt surface and the Nafion-water mixture as built by EMC were combined in LAMMPS. The interactions between the Pt atoms and other atoms in the system were obtained from Brunello, et al.[1] The Pt atoms were held fixed throughout the simulations. The thermostat was applied only to the polymer, water, and hydronium ions. The initial system was first compressed perpendicular to the Pt surfaced to $L_z = 65$ Å, and subsequent simulations applied the barostat only in the z-direction, keeping the lateral dimensions fixed throughout. Periodic boundary conditions were applied in all directions, so there are two Pt-polymer interfaces in the system.

Both the bulk and composite systems were subjected to an annealing protocol involving successive cycles of MD simulations at high temperatures and pressures in order to overcome large energy barriers. In the initial annealing runs, the atomic mass of the F atoms was set to 2.0 g/mol to help speed the polymer dynamics. Each cycle consisted of three MD runs: 1) a 50 ps NVT run at 1000 K, 2) a 100 ps NVT run at 300 K, and 3) a 50 ps NPT run at 300 K. A total of nine cycles were performed, progressively increasing the pressure to a high pressure and then decreasing back to 1 atm, using pressure values of 100, 1000, 10000, 5000, 1000, 500, 100, 10 and 1 atm. Following these cycles, the systems were equilibrated further in the NPT ensemble at 300 K and 1 atm for 2 ns.

The bulk Nafion systems were then annealed further, with an NPT run for 2 ns at 600K and 10 atm, then a ramp down to 353 K and 1 atm, at which point the F mass was reset to its correct value of 18.998 g/mol and the systems were run for another 2 ns. Finally, the systems were ramped down to 300 K and 1 atm and equilibrated for a final 1 ns. The total equilibration time for the bulk systems was approximately 8 ns. Final densities were 1.64, 1.60, and 1.59 g/cm$^3$, for the $\lambda = 9$, 12, and 15 Pt-Nafion-water systems, respectively. These densities are similar to those found in previous simulations of Nafion with the DREIDING force field and the SCP/Fw water model[2] and the densities of our Pt-Nafion-water training set (Figure S1). For the Pt-Nafion-water systems, after the annealing cycles the F mass was reset to 18.998 g/mol and the systems were further equilibrated at 300K and 1 atm for 1 ns, for a total equilibration time of approximately 3.5 ns.

## VI. ADDITIONAL RDFS, BOND LENGTH HISTOGRAMS, AND BOND ANGLE HISTOGRAMS

Plotted below are AIMD and MACE MLIP MD results for predicted RDFs, and predicted bond length and bond angle histograms. As for bond lengths, Figures S8 and S9 show that our model is also in good agreement with AIMD for bond angle predictions. Our model and the AIMD bond angle histograms have closely aligned peak positions, with our model predicting sharper histograms for C-S-O, O-C-C, O-S-O, C-C-C, C-C-S, and C-O-C bond angles, but broader histograms for H-O-H and O-S-O bond angles.

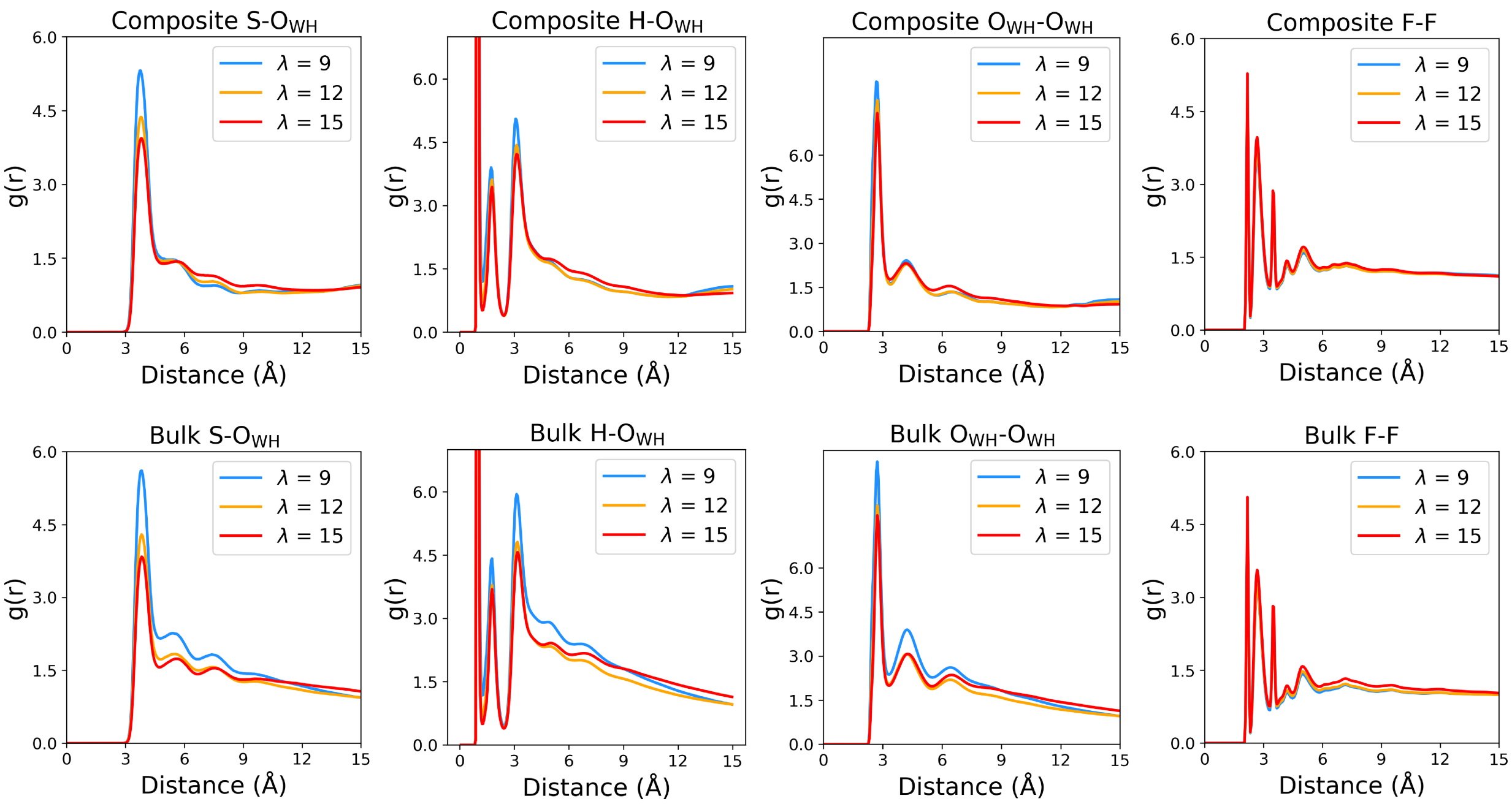

Figure S4: RDFs of bulk and composite systems for 1 ns MD trajectories generated using our trained MACE model.

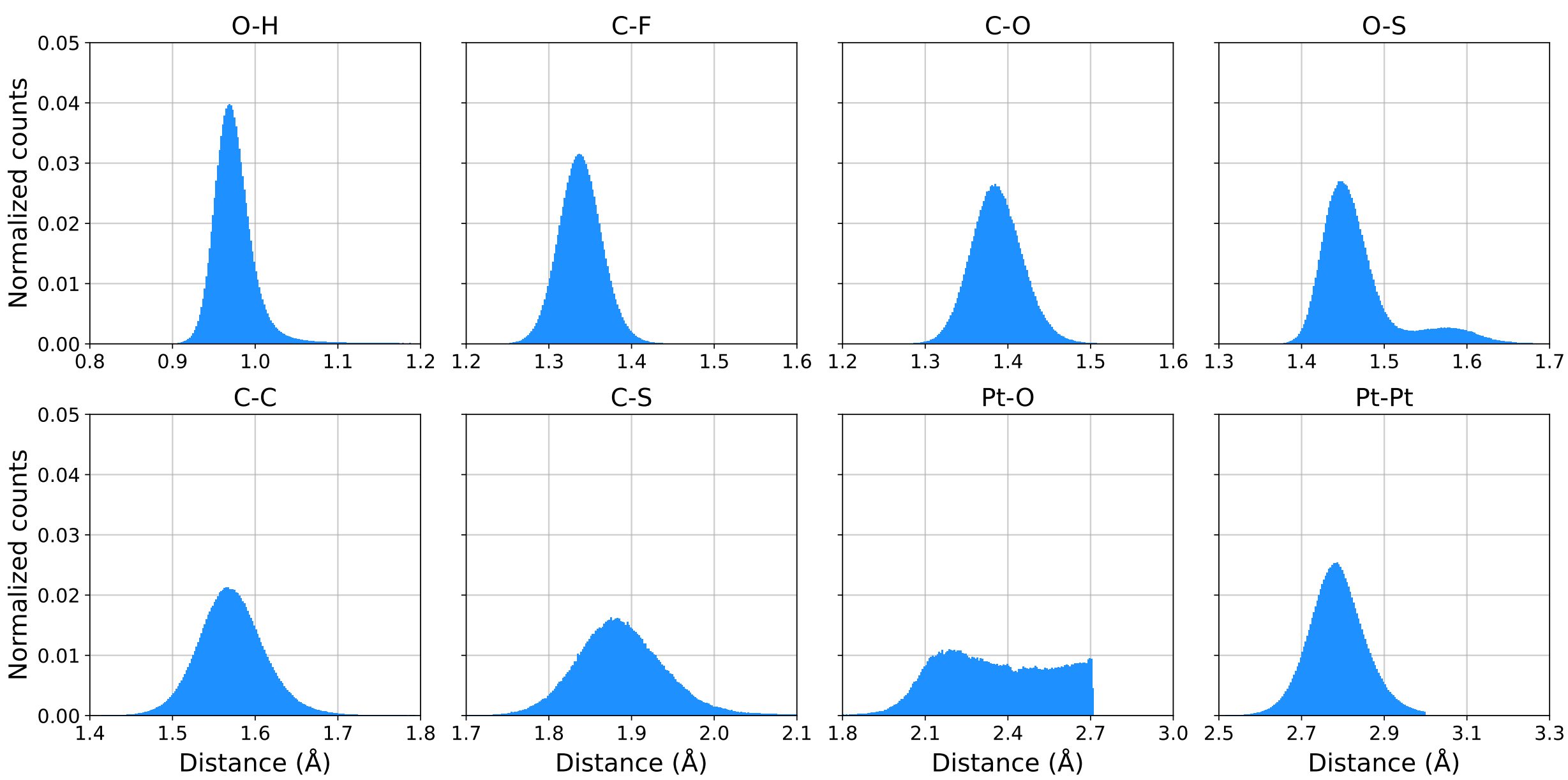

Figure S5: Bond distances of several atom type pairs extracted from the entire set of AIMD trajectories used to construct the training dataset. The subplots are sorted by mean bonding distance in ascending order. A bonded pair between atoms $A_1$ and $A_2$ was defined as occurring when the 2 atoms were closer than $\alpha(r_{A_1} + r_{A_2})$, where $r_{A_1}$ and $r_{A_2}$ are the covalent radii of $A_1$ and $A_2$. $\alpha$ is a scale factor tuned to be 1.341 so that the maximum distance for an O-H bond was 1.3 Å. For Pt-O, the scale factor results in a cutoff of 2.7 Å, resulting in a sharp distribution cutoff as oxygen atoms are present in water far from the surface. We note that the second small peak near 1.6 Å for the AIMD O-S histogram occurs when a sulfonate oxygen atom is protonated. This peak is only present for AIMD trajectories because sulfonate groups were initialized in their protonated state. In contrast, starting structures for our MLIP trajectories were initialized with deprotonated sulfonate groups and hydronium.

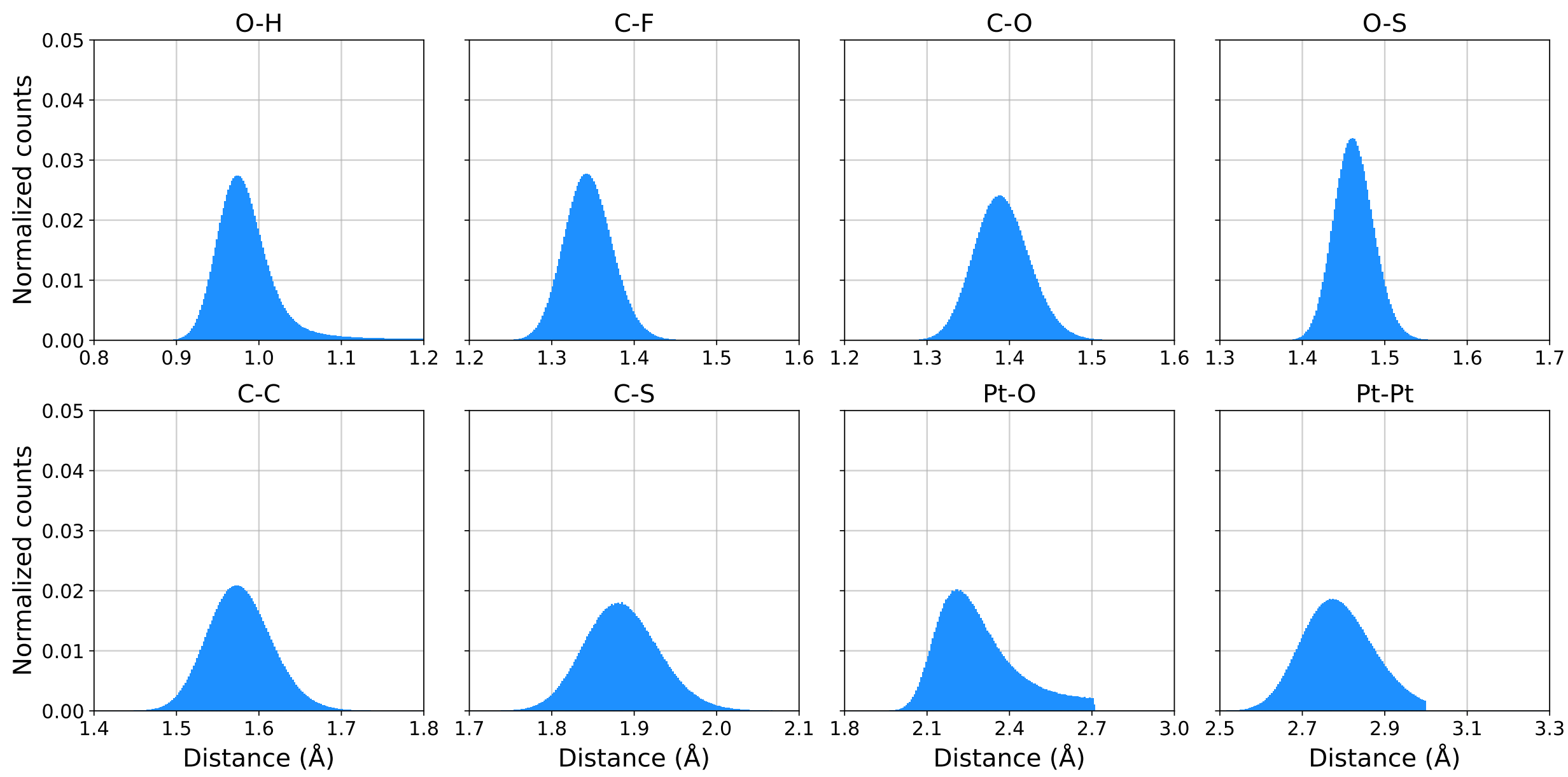

Figure S6: Bond distances of several atom type pairs across both bulk and composite systems extracted from 1 ns MD trajectories generated using our trained MACE model. The subplots are sorted by mean bonding distance in ascending order. A bonded pair between atoms $A_1$ and $A_2$ was defined as occurring when the 2 atoms were closer than $\alpha(r_{A_1} + r_{A_2})$, where $r_{A_1}$ and $r_{A_2}$ are the covalent radii of $A_1$ and $A_2$. $\alpha$ is a scale factor tuned to be 1.341 so that the maximum distance for an O-H bond was 1.3 Å. For Pt-O, the scale factor results in a cutoff of 2.7 Å, resulting in a sharp distribution cutoff as oxygen atoms are present in water far from the surface. We note that the second small peak near 1.6 Å for the AIMD O-S histogram occurs when a sulfonate oxygen atom is protonated. This peak is only present for AIMD trajectories because sulfonate groups were initialized in their protonated state. In contrast, starting structures for our MLIP trajectories were initialized with deprotonated sulfonate groups and hydronium.

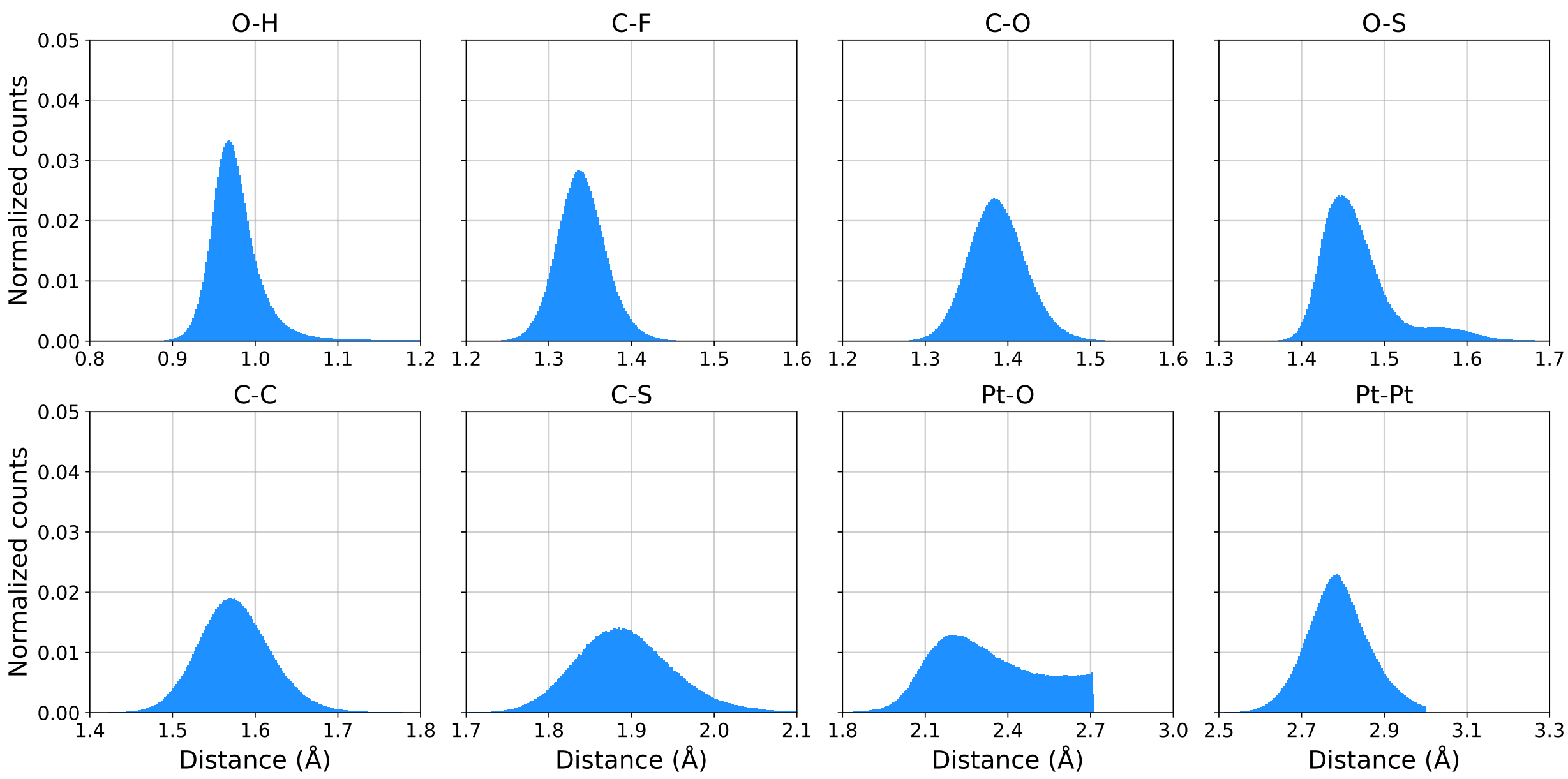

Figure S7: Bond distances of several atom type pairs extracted from 2 ps MD trajectories run on AIMD simulation cells with our trained MACE model with the same input structure as AIMD. The subplots are sorted by mean bonding distance in ascending order. A bonded pair between atoms $A_1$ and $A_2$ was defined as occurring when the 2 atoms were closer than $\alpha(r_{A_1} + r_{A_2})$, where $r_{A_1}$ and $r_{A_2}$ are the covalent radii of $A_1$ and $A_2$. $\alpha$ is a scale factor tuned to be 1.341 so that the maximum distance for an O-H bond was 1.3 Å. For Pt-O, the scale factor results in a cutoff of 2.7 Å, resulting in a sharp distribution cutoff as oxygen atoms are present in water far from the surface. We note that the second small peak near 1.6 Å for the AIMD O-S histogram occurs when a sulfonate oxygen atom is protonated. This peak is only present for AIMD trajectories because sulfonate groups were initialized in their protonated state. In contrast, starting structures for our MLIP trajectories were initialized with deprotonated sulfonate groups and hydronium.

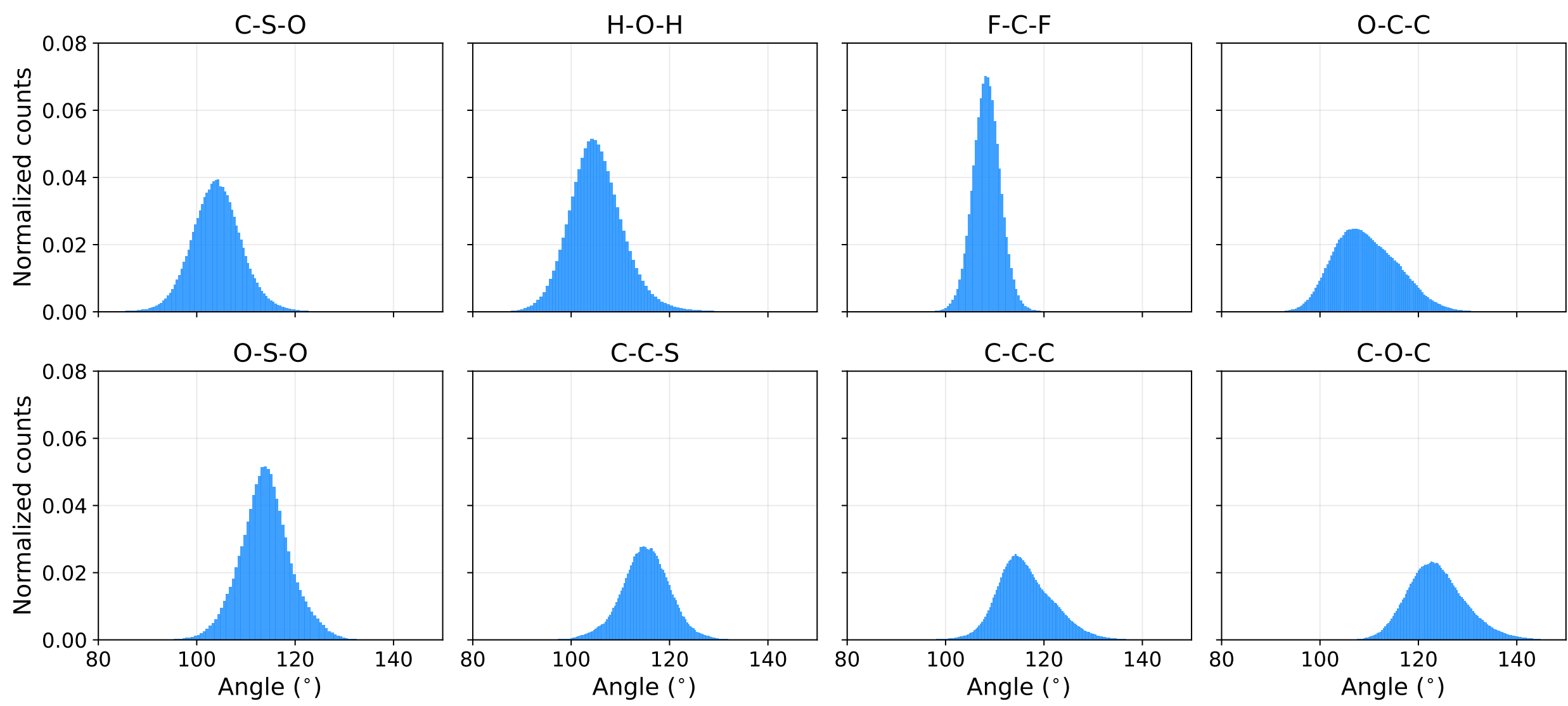

Figure S8: Bond angles of several atom type triplets extracted from the entire set of AIMD trajectories used to construct the training dataset. The subplots are sorted by mean angle magnitude in ascending order.

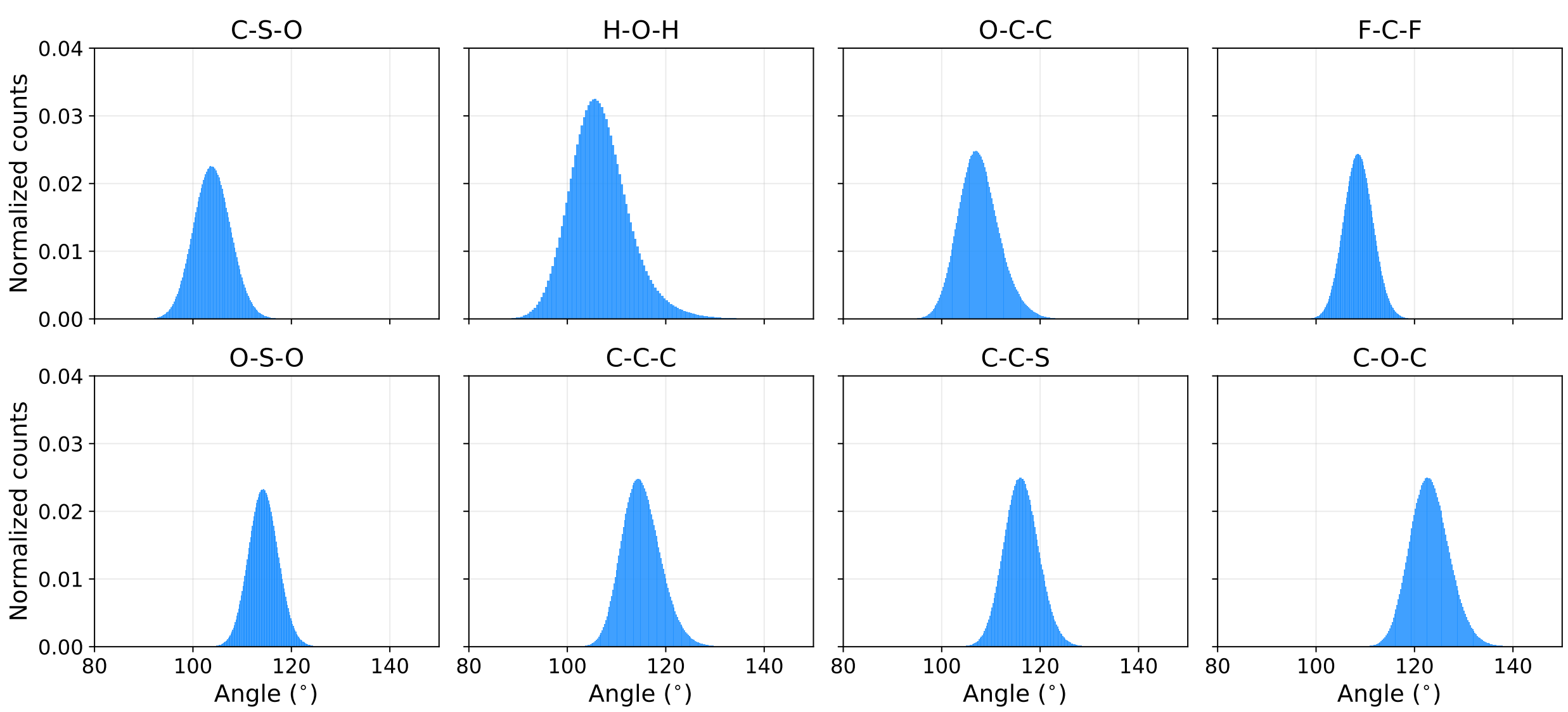

Figure S9: Bond angles of several atom type triplets across both bulk and composite systems extracted from 1 ns MD trajectories generated using our trained MACE model. The subplots are sorted by mean angle magnitude in ascending order.

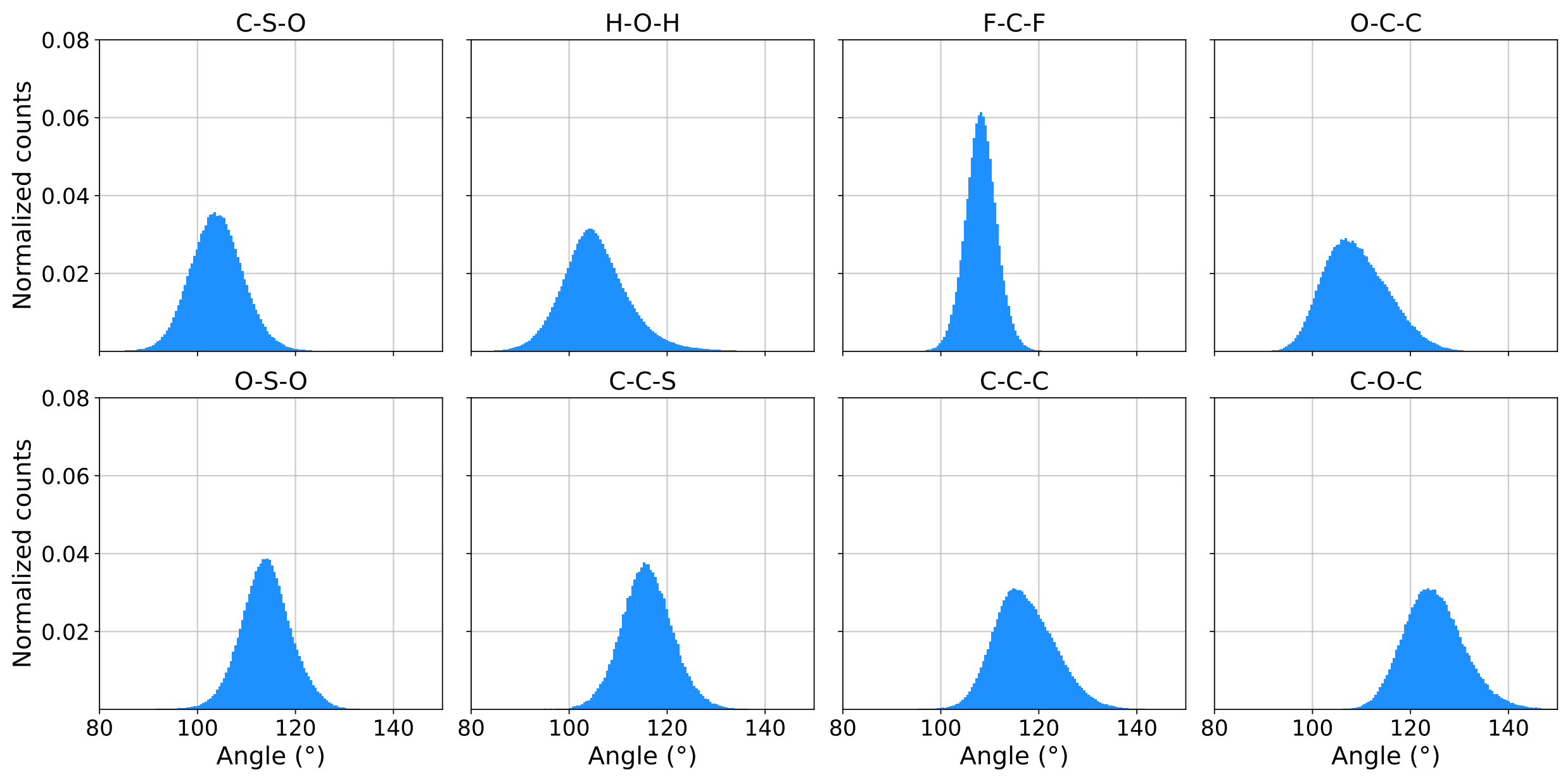

Figure S10: Bond angles of several atom type triplets across both bulk and composite systems extracted from simulations of AIMD cells using our trained model. The subplots are sorted by mean angle magnitude in ascending order.

## VII. PROTON TRACKING METHODOLOGY

The center of charge approach was developed to follow the positively charged particles of the system. In the bulk and composite systems of the present study, all positive charge starts as hydronium, and the oxygen position is tracked accordingly. At each evaluated frame in the trajectory, each positively charged particle is evaluated. If the hydronium continues as hydronium, the oxygen's coordinate is still tracked. Alternatively, if the hydronium loses a proton, the proton is tracked individually if it is free, or the newly formed hydronium if the proton hopped to a water molecule. In the case of cascades, a chain series of proton hopping events resulting in a large charge displacement, the proton defect is recursively traced to determine the positively charged particle, which is then tracked. Finally, in the case of complex formation, such as Zundel and Eigen ions, the acidic proton is determined as the shared proton for Zundel, or the smallest $H_3O$ subparticle for Eigen ions. This methodology allows the MSD of the displaced charge to be determined, gathering insight into the Grotthuss hopping mechanism. After the coordinates are gathered, the MSD is determined through a time-lagged approach utilizing autocorrelation through the Wiener–Khinchin theorem[3].

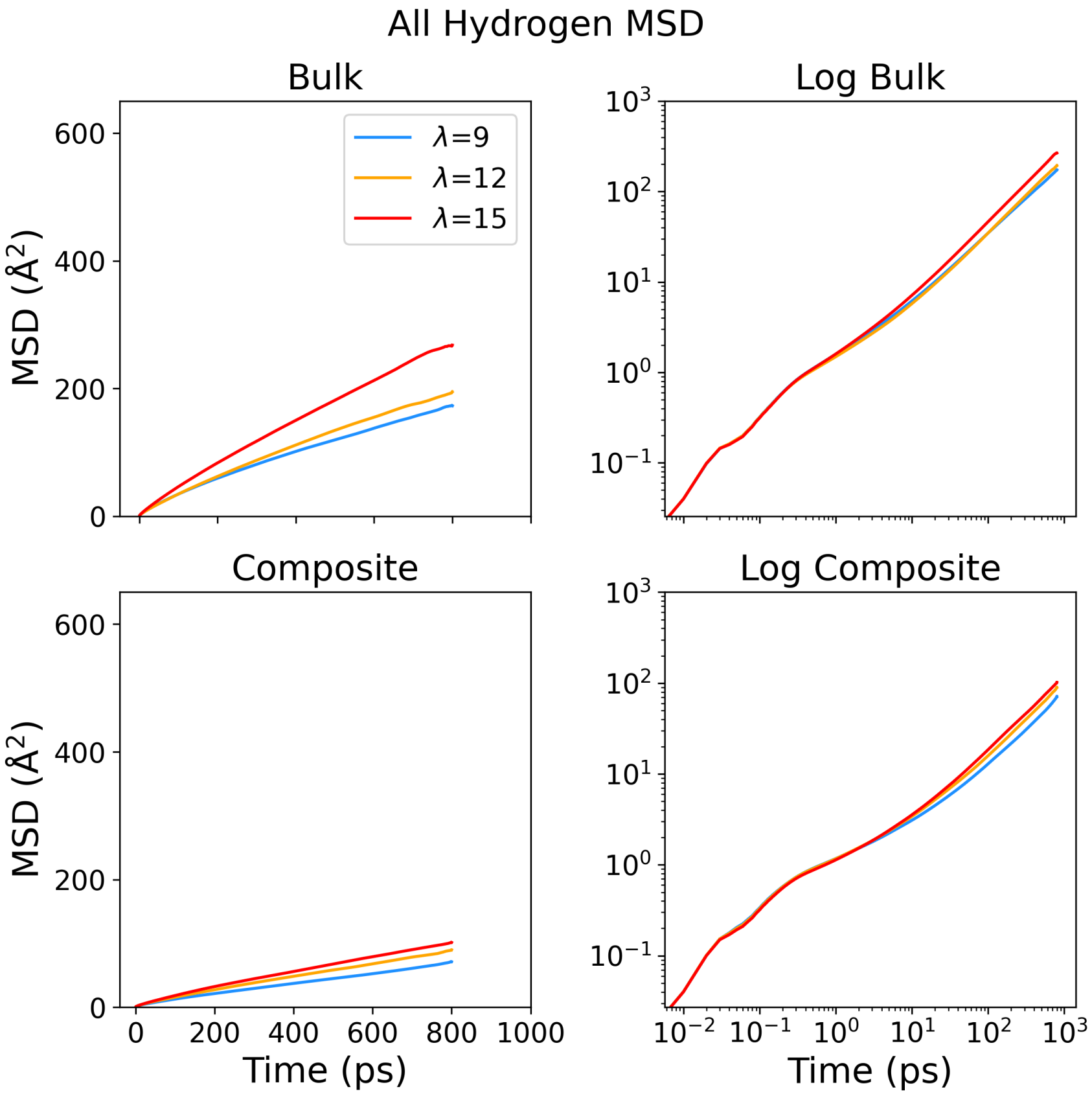

Figure S11: MSDs for all hydrogens in the system, calculated with MDAnalysis[4].

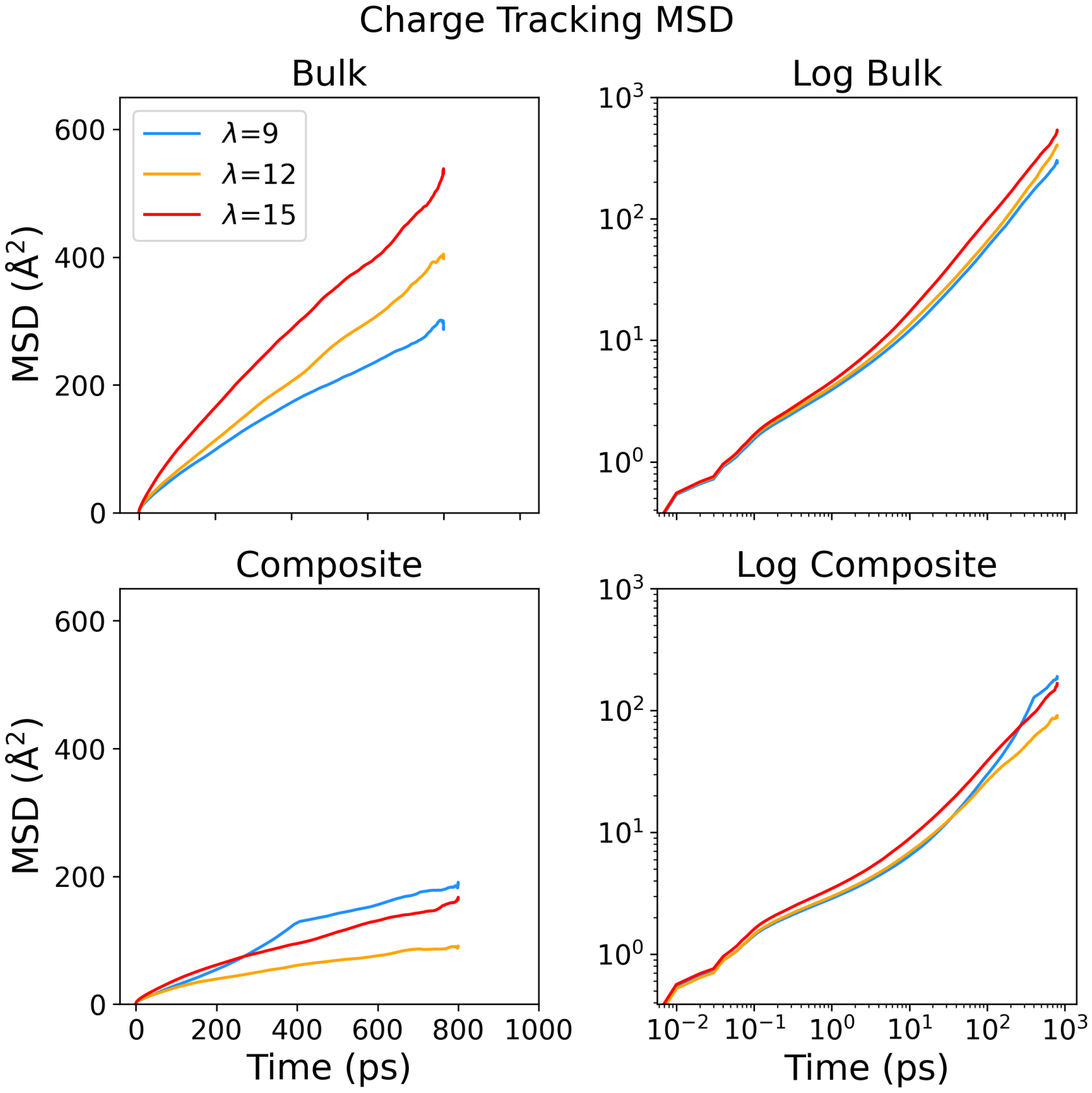

Figure S12: MSDs of proton defects calculated using the charged tracking methodology described in this section.

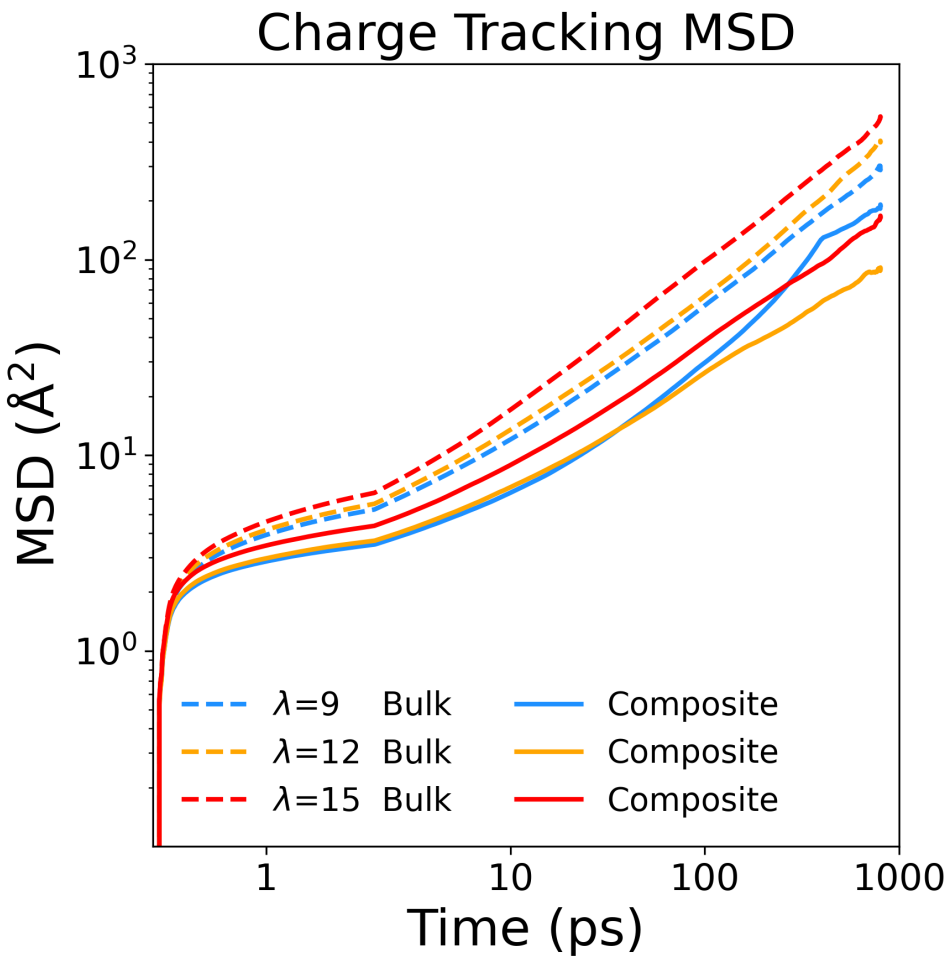

Figure S13: Log MSDs of proton defects calculated using the charged tracking methodology described in this section.

Table S6: MSDs (Å$^2$) at t = 600 ps for the MD trajectories generated using our MACE MLIP. The Bulk systems contain Nafion-water while the Composite systems contain Pt-Nafion-water.

| $\lambda$ | All Hydrogen | | Charge Tracking | |
|---|---|---|---|---|
| | Bulk | Composite | Bulk | Composite |
| 9 | 137.92 | 52.67 | 229.61 | 155.77 |
| 12 | 155.07 | 68.11 | 297.80 | 76.01 |
| 15 | 212.71 | 79.24 | 389.50 | 130.67 |

## VIII. ADDITIONAL Z-AVERAGED DENSITY PROFILES

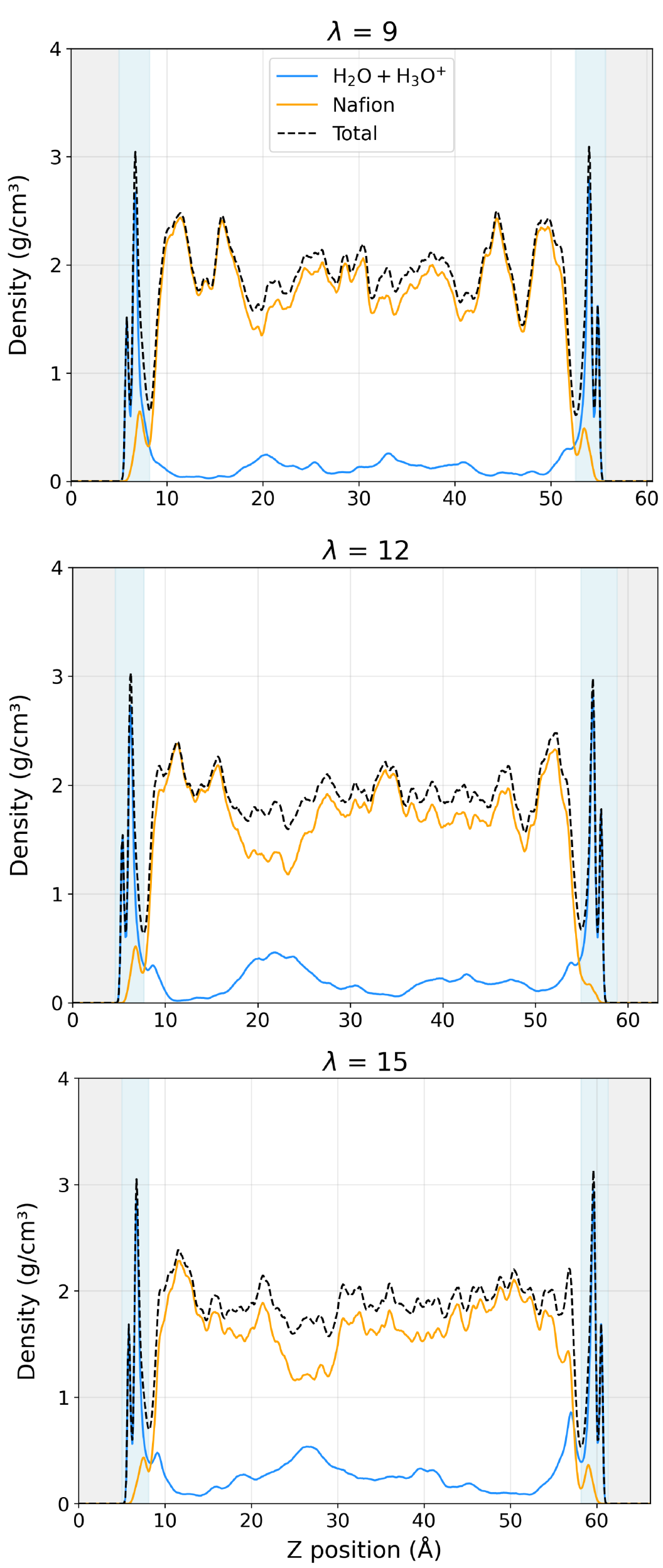

Figure S14: Planar-averaged density along the z-axis for the Pt-Nafion-water systems at 3 hydrations. The gray shaded region represents the platinum in the system, while the blue shaded region represents the surface water coordination region.

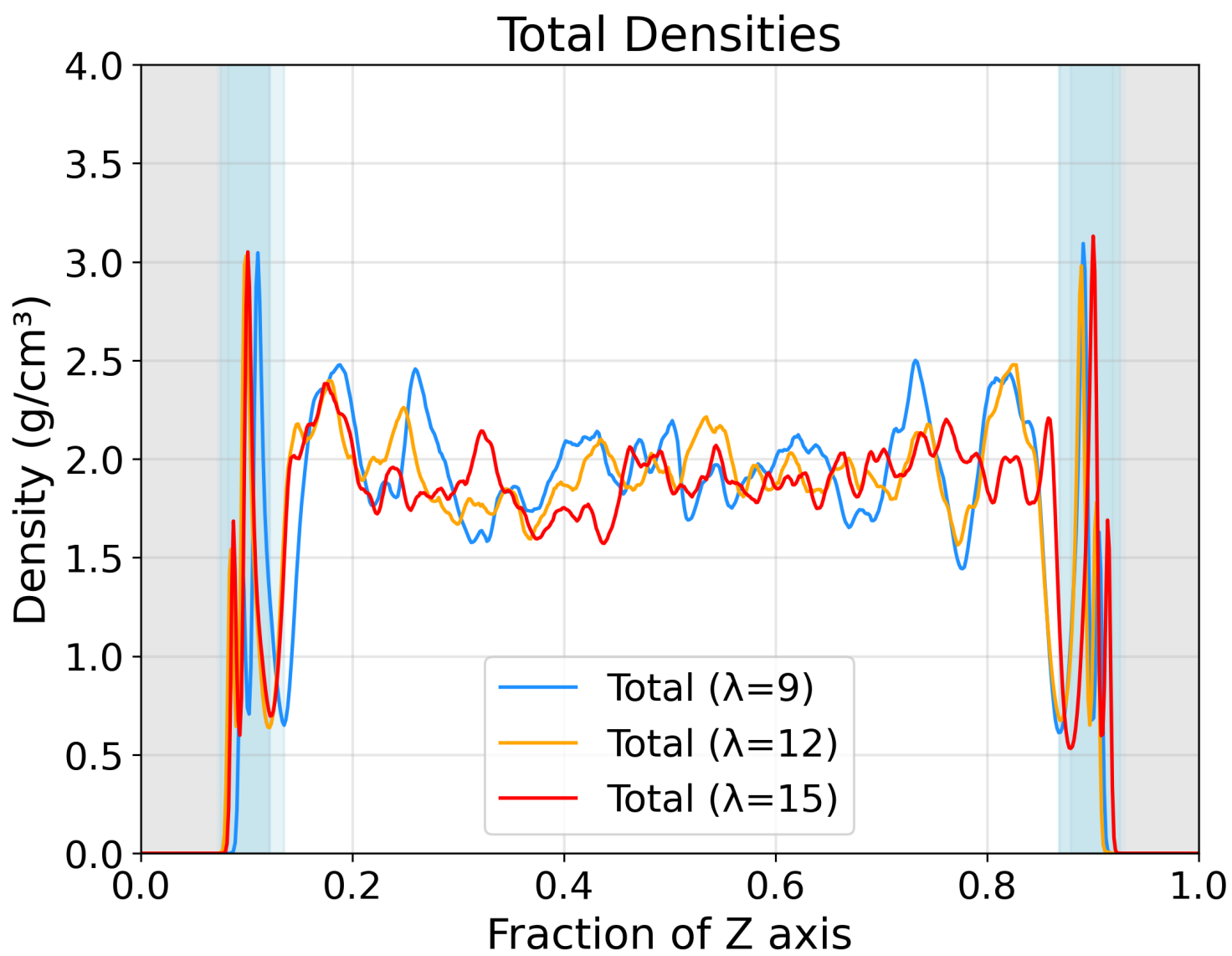

Figure S15: Total planar-averaged density along the z-axis of the Pt-Nafion-water systems at 3 hydrations. The z-axis for the 3 systems has been normalized to 1 to facilitate comparisons between hydration levels, which have different z axis lengths.

## IX. ADDITIONAL REACTION PATHWAY RESULTS

These reaction pathways were generated by selecting a single structure from the end of a Pt-Nafion-water AIMD trajectory ($\lambda$=12 with no strain applied) included in the training set to use as the reactant state, transferring atoms along a reaction coordinate according to each reaction in the set, and then using r2SCAN to relax only the atoms involved in the reaction in both the initial or product state. This isolates the impact of the reaction on each total system energy as compared to a full ionic relaxation. The reaction intermediate images were generated from the relaxed endpoint images using the image-dependent pair potential method, as implemented in the Atomic Simulation Environment software package[5]. Intermediate image MLIP, PBE, or r2SCAN energies were then calculated without further ion optimization to provide benchmark energies, meaning that the structures and energies for these elementary steps are very likely not the actual transition states for each reaction. Given that the purpose of these calculations is to assess the accuracy of the MLIP in describing stretched-bond configurations relative to the functional used to generate their training sets, exact transition state pathways are not needed. We note that some reactions tested, such as $SO_3$ and $CF_3$ group separation from the Nafion chain, did not have a transition state in the pathway, and thus required freezing the central C and S atom in the dissociated product state to prevent these groups from relaxing back to the reactant state.

These reaction pathways were generated from a $\lambda$=12 Pt-Nafion-water AIMD structure with no strain applied as the reactant state. Atoms were then transferred along each reaction coordinate above to a product state. Only the atoms involved in this reaction were then relaxed using r2SCAN to isolate the impact this each reaction on total system energy. The reaction intermediate images were generated from the relaxed endpoint images using the image-dependent pair potential method, as implemented in the Atomic Simulation Environment software package[5]. Intermediate image MLIP, PBE, or r2SCAN energies were then calculated without further ion optimization to provide benchmark energies, meaning that the structures and energies for these elementary steps are very likely not the actual transition states for each reaction. Given that the purpose of these calculations is to assess the accuracy of the MLIP in describing stretched-bond configurations relative to the functional used to generate their training sets, exact transition state pathways are not needed.

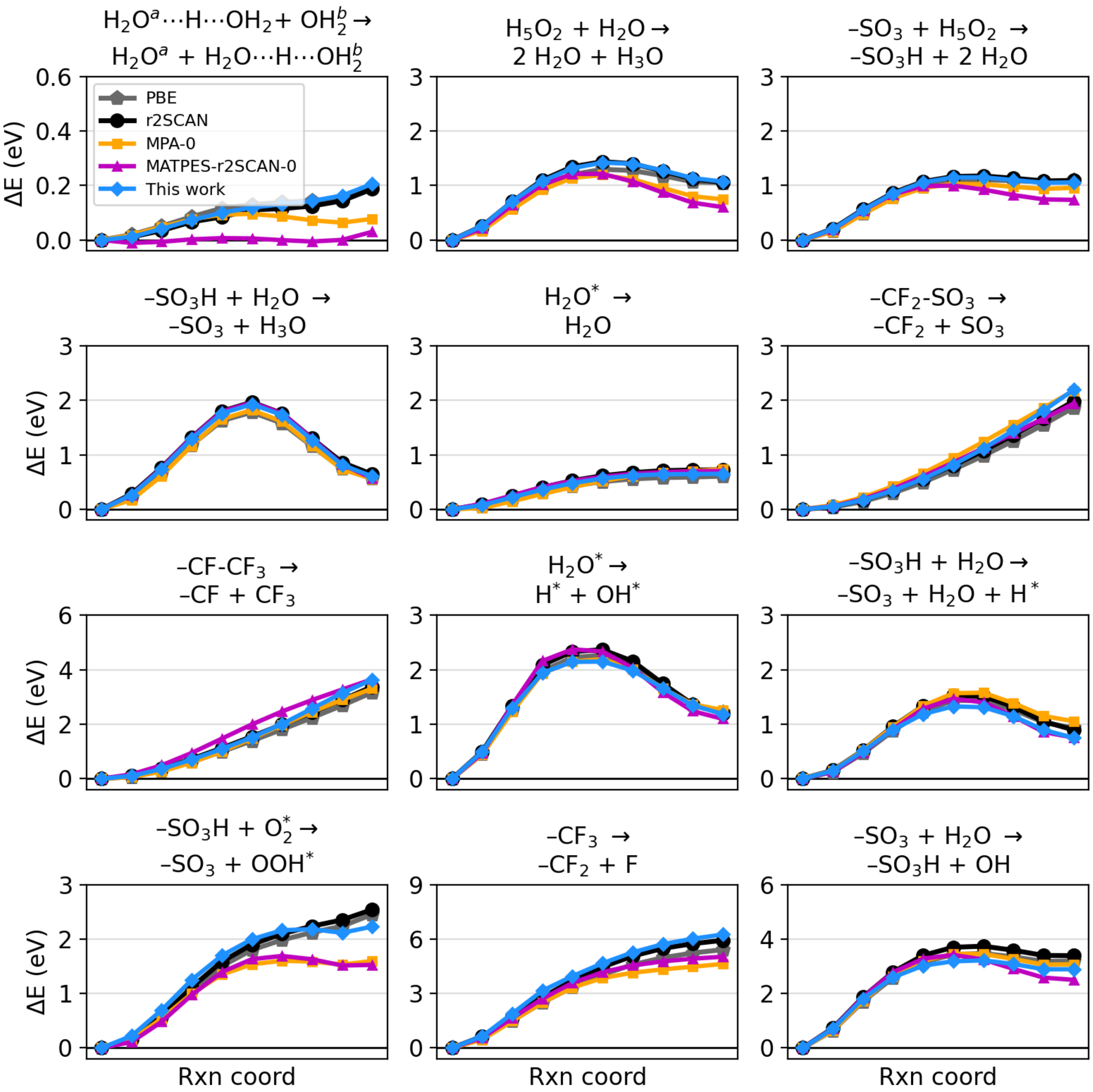

Figure S16: Reaction pathway energies predicted by the MLIP model trained in this work compared to those predicted by the MACE-MATPES-r2SCAN-0 medium model, MACE-MPA-0 medium model, and PBE and r2SCAN DFT calculations. We note that these pathways are single point energies for targeted atom transfers and are very likely not the true transition state energies for these elementary steps.

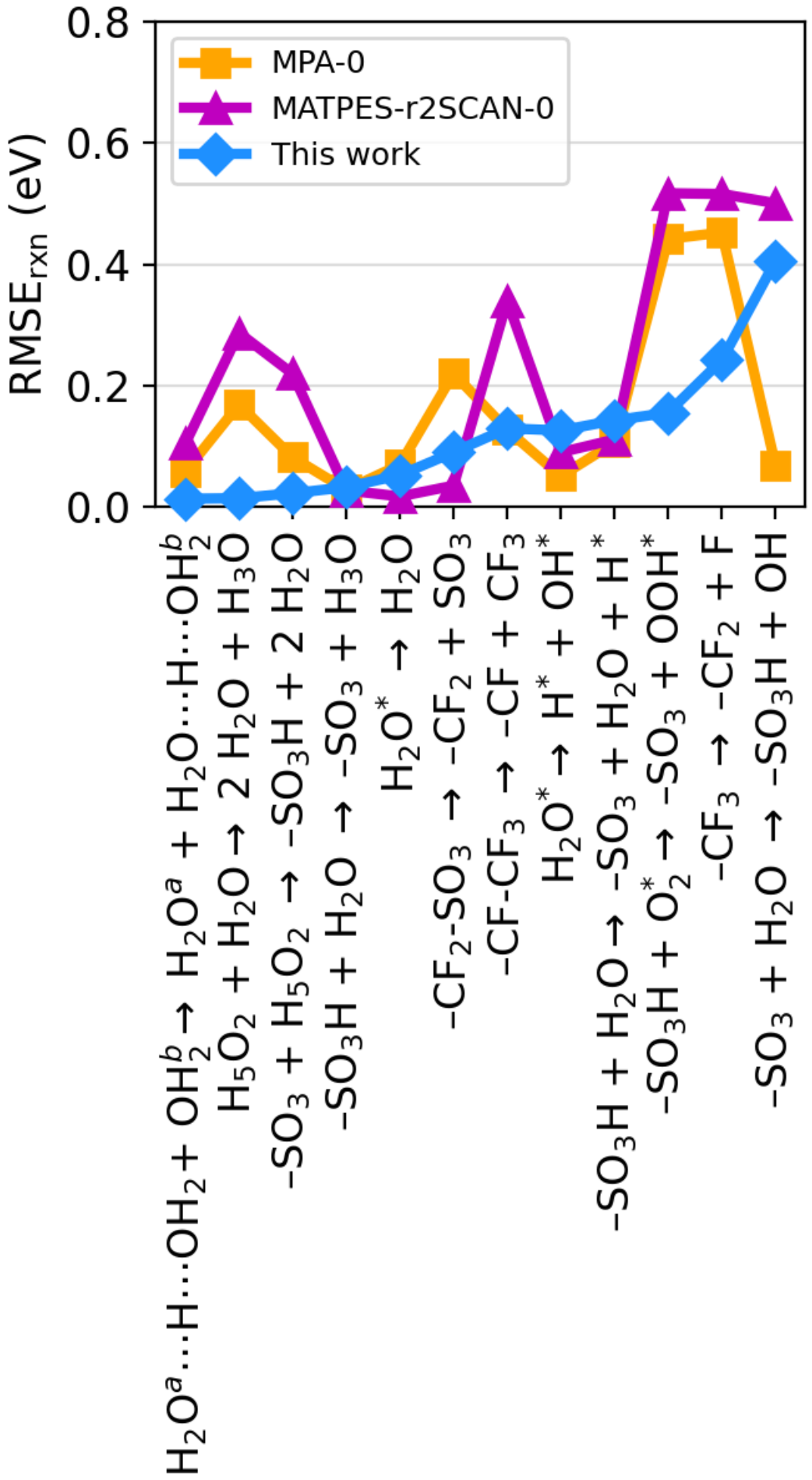

Figure S17: Reaction pathway energy root mean squared error (RMSE) averaged over all 10 images in each reaction pathway for the MLIP model trained in this work compared to the MACE-MATPES-r2SCAN-0 medium model, MACE-MPA-0 medium model, and r2SCAN DFT. The starting state (image 0) for each reaction was excluded from the RMSE calculations shown in this plot because its energy was always shifted to be 0 eV. We note that these pathways are single point energies for targeted atom transfers and are very likely not the true transition state energies for these elementary steps.

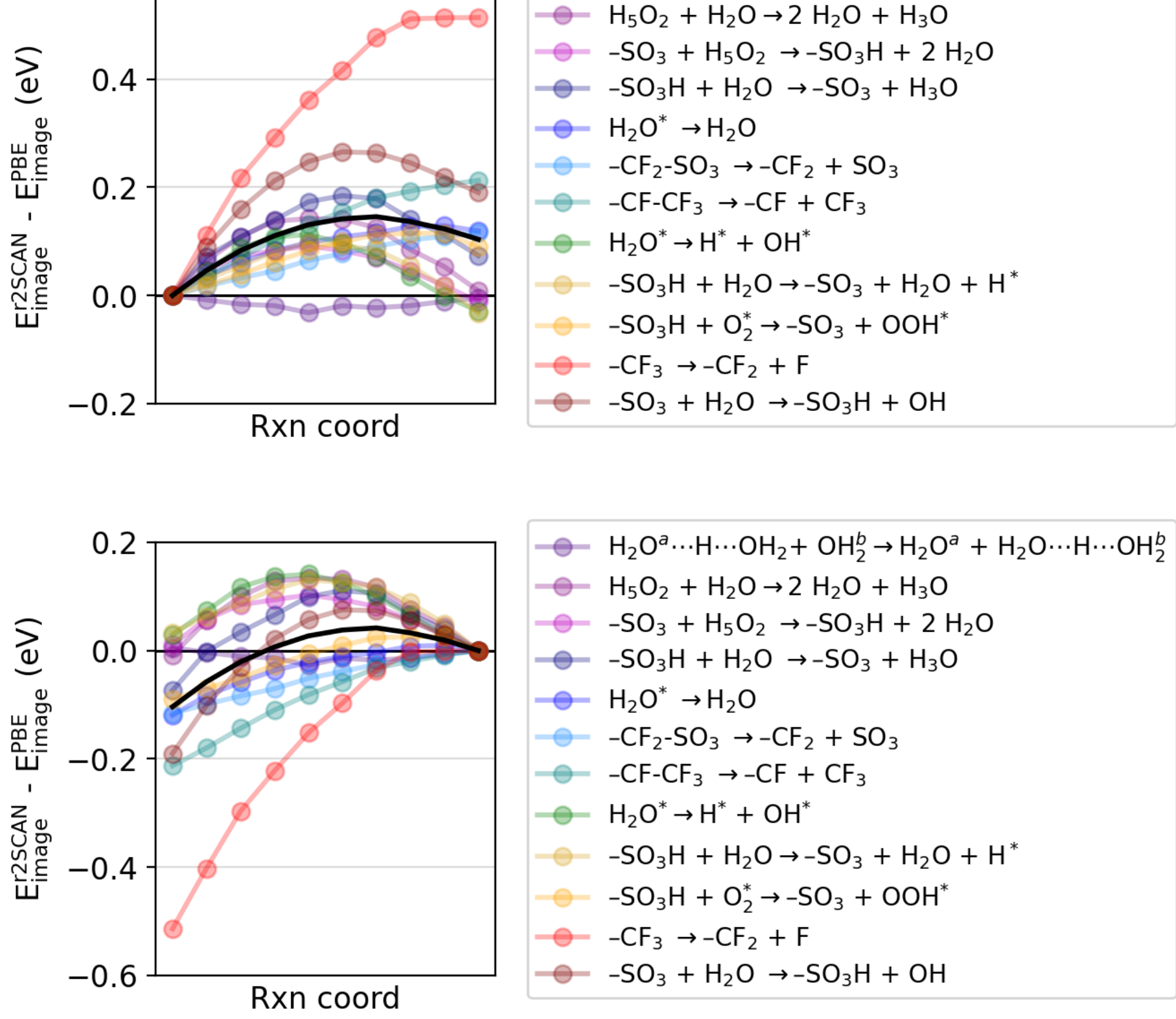

Figure S18: Reaction pathway energy differences for PBE vs r2SCAN. The average energy difference across all reactions at the same image index is shown as the solid black data. As the reaction coordinates of these pathways are arbitrary, the starting state (image 0) for each reaction was always shifted to be 0 eV in the top subplot, while the ending state (image 9) for each reaction was always shifted to be 0 eV in the bottom subplot. We note that these pathways are single point energies for targeted atom transfers and are very likely not the true transition state energies for these elementary steps.